\documentclass{aa}  
\usepackage{graphicx}
\usepackage{txfonts}

\usepackage{bm}
\usepackage{color}
\usepackage{url}
\usepackage{enumitem}

\usepackage{hyperref}
\usepackage{balance}
\begin{document} 

\def\be{\begin{equation}}
\def\ee{\end{equation}}
\def\d{\mbox{\rm d}}
\def\LCDM{$\Lambda$CDM\,}

\title{MOND-like behavior in the Dirac-Milne universe}
\subtitle{Flat rotation curves and mass/velocity relations in galaxies and clusters}

\author{Gabriel Chardin
	\inst{1}
	\and
	Yohan Dubois\inst{2}
	\and
	Giovanni Manfredi\inst{3}
	\and
	Bruce Miller\inst{4}
	\and
	Cl\'ement Stahl\inst{1}
	}

\institute{Universit\'e de Paris, CNRS, Astroparticule et Cosmologie,  F-75006 Paris, France\\
\href{mailto:chardin@apc.in2p3.fr}{email: chardin@apc.in2p3.fr}
\and
Institut d'Astrophysique de Paris, UMR 7095, CNRS \& Sorbonne Universit\'e, 98 bis Boulevard Arago, F-75014 Paris, France
\and
Universit\'e de Strasbourg, CNRS, Institut de Physique et Chimie des Mat\'eriaux de Strasbourg, UMR 7504, F-67000 Strasbourg, France\\
\href{mailto:giovanni.manfredi@ipcms.unistra.fr}{email: giovanni.manfredi@ipcms.unistra.fr}
\and
Department of Physics and Astronomy, Texas Christian University, Fort Worth, TX 76129, USA
}
\date{\today}


\abstract
{}
{Observational data show that the observed luminous matter is not sufficient to explain several features of the present universe,
from gravitational structure formation to the rotational velocities in galaxies and clusters. The mainstream
explanation is that the missing mass, although gravitationally active, interacts very weakly with ordinary matter. Competing explanations
involve changing the laws of gravity at low accelerations, as in MOND (Modified Newtonian Dynamics). Here, we suggest that the
Dirac-Milne cosmology, a matter-antimatter symmetric cosmology where the two components repel each other,
is capable of accounting for such apparent modification of the Newtonian law, without invoking dark matter.}
{Using a simple analytical approximation and 1D and 3D simulations, we study rotation curves and virial velocities, and compare the mass observed in the simulations to the mass derived assuming Newtonian gravity. Using a modified version of the \texttt{RAMSES} code, we study the Faber-Jackson scaling relation
and the intensity of the additional gravitational field created by antimatter clouds.
}
{We show that, in the Dirac-Milne universe, rotation curves are generically flat beyond a characteristic distance of $\approx 2.5$ virial radii
, and that the Tully-Fisher and Faber-Jackson scaling relations with exponent $\approx 3$ are satisfied. 
We show that the mass derived from the rotation curves assuming Newtonian gravity is systematically overestimated compared to
the mass really present. In addition, the Dirac-Milne universe, featuring a polarization between its matter and
antimatter components, presents a behavior similar to that of MOND,
characterized by an additional surface gravity compared to the Newtonian case.
We show that in the Dirac-Milne universe, at the present epoch, the intensity of the additional gravitational field $g_{am}$ due
to the presence of clouds of antimatter is of the order of a few $10^{-11}$ m/s$^2$,
similar to the characteristic acceleration of MOND. We study the evolution of this additional acceleration $g_{am}$
and show that it depends on the redshift, and is therefore not a fundamental constant.}
{Combined with its known concordance properties on SNIa luminosity distance, age, nucleosynthesis
and structure formation, the Dirac-Milne cosmology may then represent an interesting alternative
to the \LCDM, MOND, and other scenarios for explaining the dark matter (or missing gravity) and dark energy conundrum.}

\keywords{dark matter --
                dark energy --
                cosmology --
		gravitation               }
\maketitle

\section{Introduction}\label{sec:intro}

The dark matter enigma found its first expression in the 1930s after the observation by Fritz Zwicky \citep{Zwicky1933} that peculiar velocities in the Coma cluster were far too large (by more than two orders of magnitude according to Zwicky's analysis) to account for the bound behavior of the cluster components if only the visible mass was taken into account. Following this initial and remarkably prescient observation and analysis, a long dormant period followed where the dark matter question was mostly forgotten
 \citep[however, see][]{Kahn_Woltjer_1959}. In the 1970s, Vera Rubin and collaborators \citep{Rubin1980}, and Albert Bosma \citep{Bosma_1981}, measuring galactic rotation curves, noticed that they had quite generally a flat behavior at large distances from the core, and their analysis contributed very significantly to the revival of the dark matter enigma. The accumulation of galactic rotation curves then led to the gradual realization \citep{Bertone_Hooper_2018} that, quite generally, observed galaxy rotation curves are flat at large distances from the galaxy core, which is at odds with theoretical predictions based on the assumption of dominating mass related to luminous matter.

Two main lines of hypotheses were proposed as tentative solutions to this enigma:
\begin{itemize}
\item The first was to conjecture that there really exists a large part of the matter component of our Universe that is  dark and interacts very weakly, apart from its gravitational interactions. This conjecture was for a long time the dominant hypothesis, under the implementation of WIMPs (Weakly Interacting Massive Particles), and later CDM (cold dark matter), after the findings  that both massive neutrinos \citep{Drukier_Stodolsky_1984} and  supersymmetric particles \citep{Goodman_Witten_1984}  provided candidates that naturally suited the existing constraints, in terms of mass and interaction cross-sections, to solve the dark matter problem -- the so-called ``WIMP miracle''. This hypothesis seemed indeed almost necessary since the then mainstream Einstein-de-Sitter (EdS) universe featured a critical density: dark matter suggested an elegant way to fill the gap between the small baryonic component, with a density derived and constrained by nucleosynthesis to less than 5\% of the critical density, and the critical density of the EdS universe.

But despite extensive experimental searches, both in direct detection experiments and at the Large Hadron Collider at CERN, no such dark matter candidates have been found.  Meanwhile, the tensions between the age of an Einstein-de Sitter  and  the age of the oldest structures in the Universe \citep{Bolte_Hogan_1995} grew to a point in the mid-1990s where a cosmological constant or some other repulsive component was considered compulsory, a few years before the discovery in 1998 of what is now called dark energy, through the SNIa {flux} measurements \citep{Riess98,Perlmutter99}.

The standard cosmological model, \LCDM, although an impressive fit to diverse sets of data, is not without presenting some tensions: probably the highest tension originates from the local measurements of the Hubble constant $H_0$ on the one hand, and the determinations of this same parameter deduced from the cosmic microwave background (CMB) and baryonic acoustic oscillations (BAO) on the other\,\citep{Riess2020,DiValentino_2021}. But even on the dark matter side, additional tensions exist: there is for example no evidence for the cusped density profiles predicted by the dark matter simulations\,\citep{Flores_1994}, and many galaxies apparently have no need for dark matter in their central regions. For some galaxies, there is even no apparent need for dark matter altogether, requiring to explain how the dominant dark matter component may have been almost entirely ejected from these galaxies \citep[see \emph{e.g.}][and references therein]{Guo_DM_less_dwarf_galaxies_2020}.
This scenario also suffers from the fact that its two main ingredients, dark energy and dark matter, supposed to represent 95\% of its content, have not been identified despite extensive investigations. The extremely rare coplanarity of the satellites around
Andromeda\,\citep{Ibata_Nature_2013},  Centaurus A\,\citep{Muller_Science_2018}, and the Milky Way\,\citep{Pawloswski_Kroupa_2020}, at present the only three galaxies with well-known satellite distributions, is an indication of additional tension in the \LCDM scenario. Finally, we note  the too-big-to-fail problem\,\citep{Boylan-Kolchin_2011}, possibly addressed by baryonic back-reaction \citep{Governato_2012, Teyssier_2013, Wetzel_2016}.

\item The second line of explanation rests on a modification of the laws of gravitation,  MOdified Newtonian Dynamics less
 \citep[MOND;][]{Milgrom1983} being its most popular expression. The MOND hypothesis was initially phenomenological, and based on the observation that the dark matter problem seemed mostly confined to regions with values of the gravitational field $\la 10^{-10}$ m/s$^2$. Indeed, MOND proposes that the law of gravitation deviates from its Newtonian expression in the following way:
$a = \sqrt{a_0 a_N}$
where $a$ is the acceleration, $a_0$ is the crossover acceleration of MOND, and $a_N$ is the acceleration predicted by Newton's law.
With this single hypothesis, MOND predicted the Tully-Fisher relation \citep{Tully_Fisher_1977} linking the mass to the rotation velocity for structures over a wide range of mass (at least four orders of magnitude). And although the MOND crossover acceleration, $a_0$, seems to differ when the analysis is done at galactic or at cluster scales, a reappraisal of the systematic errors in the cluster mass profiles\,\citep{Ettori_clusters_2019}, and hybrid models with neutrinos as hot dark matter\,\citep{Angus_2009,Haslbauer_2020} could alleviate this tension.
Looking for a field-theoretical expression, initial versions proposed for MOND, such as TeVeS \citep{Bekenstein_TeVeS}, were recently ruled out by the multi-messenger observations of GW170817, but a slightly modified form passes this test\,\citep{Skordis_Zlosnik_2019}. Possible links between MOND and cosmology include the entropic expression of gravity\,\citep{Pazy_2013}, emergent gravity\,\citep{Verlinde_2017}, mimetic gravity\,\citep{Vagnozzi_2017} or some expressions of quantum gravity\,\citep{Smolin_2017}. The fascinating proximity between the values of $a_0$ and $c H_0$ is a suggestion of yet another link between cosmology and this crossover acceleration. On the other hand, we will see in the following that Dirac-Milne may provide an explanation for this coincidence. A fundamental remark to which we will come back is the fact that MOND's law looks akin to the effect of gravitational polarization\,\citep{Blanchet_2007,Blanchet_LeTiec_2009}.
\end{itemize}

In the present paper, we study the gravitational polarization predicted by the Dirac-Milne (D-M) cosmology \citep{Benoitlevy, Chardin2018}, providing an explanation for this apparent modification of the Newtonian law of gravitation. For this purpose, in Sec.\,\ref{sec:diracmilne}, we recall the main characteristics of the D-M universe. In Sec.\,\ref{sec:DM_rotations_curves}, we study a simple idealized analytical model, showing that in the D-M cosmology rotation curves are indeed expected to be generically flat after a characteristic distance, for which we provide an approximate relation. We discuss the law obtained for the rotation velocity, notably in relation to the shell and Birkhoff theorems\,\citep{Newton_1760, Birkhoff_1923}, with a more detailed analysis in Appendix\,\ref{app:A}. In Sec.\ref{sec:simulations}, we present preliminary results of velocity distributions as a function of mass using a modified version of the \texttt{RAMSES} 3D simulation code incorporating the gravitational behavior of the D-M universe, as described in \citet{Manfredi2018} and \citet{Manfredi2020}. In Sec.\,\ref{sec:TFR}, we show that D-M follows a Faber-Jackson relation with a very small scatter and an exponent $\approx 3$ over a range of more than three orders of magnitude in mass. In Sec.\,\ref{sec:MOND}, we discuss in more detail the MOND-like behavior of D-M and in particular the value of the predicted acceleration parameter $a_0$. In the final section, we summarize our findings, and provide some perspectives and possible lines of development for future work.

\section{The Dirac-Milne universe}\label{sec:diracmilne}
The Dirac-Milne (D-M) universe, proposed recently \citep{Benoitlevy, Chardin2018}, suggests a radically different paradigm for the cosmology of our universe. It features a symmetric matter-antimatter universe, where matter and antimatter effectively repel each other, but where antimatter also repels itself. As is well-known, when the usual expression of the Weak Equivalence Principle is respected for matter and antimatter, such symmetric matter-antimatter cosmologies are excluded by the non-observation of a diffuse gamma-ray flux \citep{Omnes1972, Cohen1998}. On the other hand, the D-M cosmology is a gravitational implementation of the Dirac particle-hole sea, analogous to the electron-hole system in a semiconductor. Given the fact that a repulsive and enigmatic dark energy component represents $\approx 70 \%$ of the energy density in the \LCDM model, it seems interesting to test more extensively this hypothesis of repulsion between matter and antimatter. The D-M cosmology is further motivated by the remark by \citet{Price_ajp1993} that the usual expression of the Weak Equivalence Principle, stating that all particles must follow the same trajectories given the same initial conditions in a gravitational field (universality of free fall), must necessarily be modified in the case where particles with negative mass are considered along with particles of positive mass.
Indeed, as shown by Price, two new elements appear in such bound systems: (i) Gravitational polarization appears between particles of positive and negative mass whenever they are bound by non-gravitational forces, and (ii) levitation is predicted for a symmetric $(+m,-m)$ system, a gross violation of the usual formulation of the Weak Equivalence Principle.

A fundamental feature of the D-M universe is that its expansion factor varies linearly with time:
\begin{equation}\label{linear_expansion}
a(t) \propto t
\end{equation}
while in D-M there is neither dark matter, nor dark energy  beyond its matter and antimatter components. Being a universe that appears gravitationally empty at large scales,
the initial phases of the D-M universe have timescales differing rather dramatically from the \LCDM universe: for example, the Quark-Gluon-Plasma transition lasts for about
one day, instead of about 10 microseconds in the Standard Model, while nucleosynthesis lasts about 35 years, compared to three minutes in the Standard Model, and
recombination occurs at an age of about 14 million years, compared to the 380 000 years of the \LCDM model\,\citep{Benoitlevy, Chardin2018}.

Despite these tremendous differences in the initial timescales, the D-M universe, with only one adjustable parameter $H_0$, presents several elements of concordance: its age, equal to $1/H_0$, is almost equal to the age of the \LCDM universe for $H_0 \approx 70$ km s$^{-1}$ Mpc$^{-1}$, while the $H(z)$ dependence of cosmic chronometers is nicely reproduced in coasting universes \citep{Melia_Cosmic_Chrono_2013}, along with primordial nucleosynthesis\,\footnotemark[1] and the SNIa luminosity distance \citep{Sethi_1999, Chodorowski_2005, Benoitlevy}. Also, its non-linear structure formation mechanism appears to reproduce the main features of the matter power spectrum starting from a single scale of matter-antimatter domains at decoupling \citep{Manfredi2018, Manfredi2020}.
\footnotetext[1]{It has been noted by \cite{Lewis_Barnes_2016} that coasting universes would destroy almost completely deuterium and helium-3. In the D-M cosmology, however, secondary annihilation processes produce deuterium (and helium-3), although the precise amount remains to be determined. In addition, the very slow ``simmering'' thermal nucleosynthesis, lasting $\approx35$ years, produces small quantities of elements heavier than $^4$He, $^7$B and $^7$Li, notably nitrogen and carbon nuclei. It should be noted, on the other hand, that $^7$Li is only produced in D-M at the level of $^7$Li/H\,$\approx 3 \times 10^{-10}$, alleviating in part the so-called ``lithium problem'' met by \LCDM.}
In addition, the D-M cosmology does not suffer from the horizon problem \citep{Benoitlevy} and therefore does not require inflation.

\section{Gravitational setup in the D-M universe}\label{sec:DM_rotations_curves}
It is fundamental to note that although existing in Nature, the Dirac particle-hole system has no Newtonian expression, even when the three Newtonian mass parameters (inertial, gravitational active, gravitational passive) are used \citep{Manfredi2018}.
On the other hand, as studied in \citet{Manfredi2018, Manfredi2020}, the gravitational sector of matter and antimatter in the D-M universe can be expressed with two separate gravitational potentials, using the following coupled Poisson equations:
\begin{eqnarray}
\nabla^2\phi_{+} &=& 4\pi G (\rho_{+} - \rho_{-}), \label{eq:poiss_diracmilne1}\\
\nabla^2\phi_{-} &=& 4\pi G (-\rho_{+} - \rho_{-}) \label{eq:poiss_diracmilne2} .
\end{eqnarray}

It should be noted that in these equations, although two potentials $\phi_{+}$ and $ \phi_{-}$ are invoked, there is a {\it single} gravitational constant $G$, and not two independent constants for matter and antimatter. In particular, following \cite{Price_ajp1993}, we can derive the gravitational field for antimatter once the gravitational field for matter is known (and vice-versa). For particles with equal but opposite mass, the gravitational fields:
{ 
\be
\label{eq:g+}
\vec{g}_+\equiv - \vec{\nabla} \phi_+
\ee and 
\be \vec{g}_-\equiv - \vec{\nabla} \phi_-
\ee
} 
exerted on a particle and its antiparticle are opposite in the ``Newtonian" regime (\emph{i.e.} when the gravitational field created by matter is much larger than the contribution of antimatter, which can then be neglected), as the total force on the bound  $(+m,-m)$ system is zero\,\citep{Manfredi2018, Manfredi2020}.

For further use, we also introduce the gravitational field acting on a particle sourced only by matter as:
\be
\label{eq:gmatter}
\vec{g}_m\equiv - \vec{\nabla} (\phi_+ - \phi_-)/2
\ee
and the corresponding field acting on a particle sourced only by antimatter:
\be
\label{eq:gantimatter}
\vec{g}_{am}\equiv - \vec{\nabla} (\phi_+ + \phi_-)/2
\ee

\subsection{General properties of matter, antimatter and depletion zone}
\label{sec:repart}

\begin{figure}[]
\begin{center}
\includegraphics[width=0.8\linewidth]{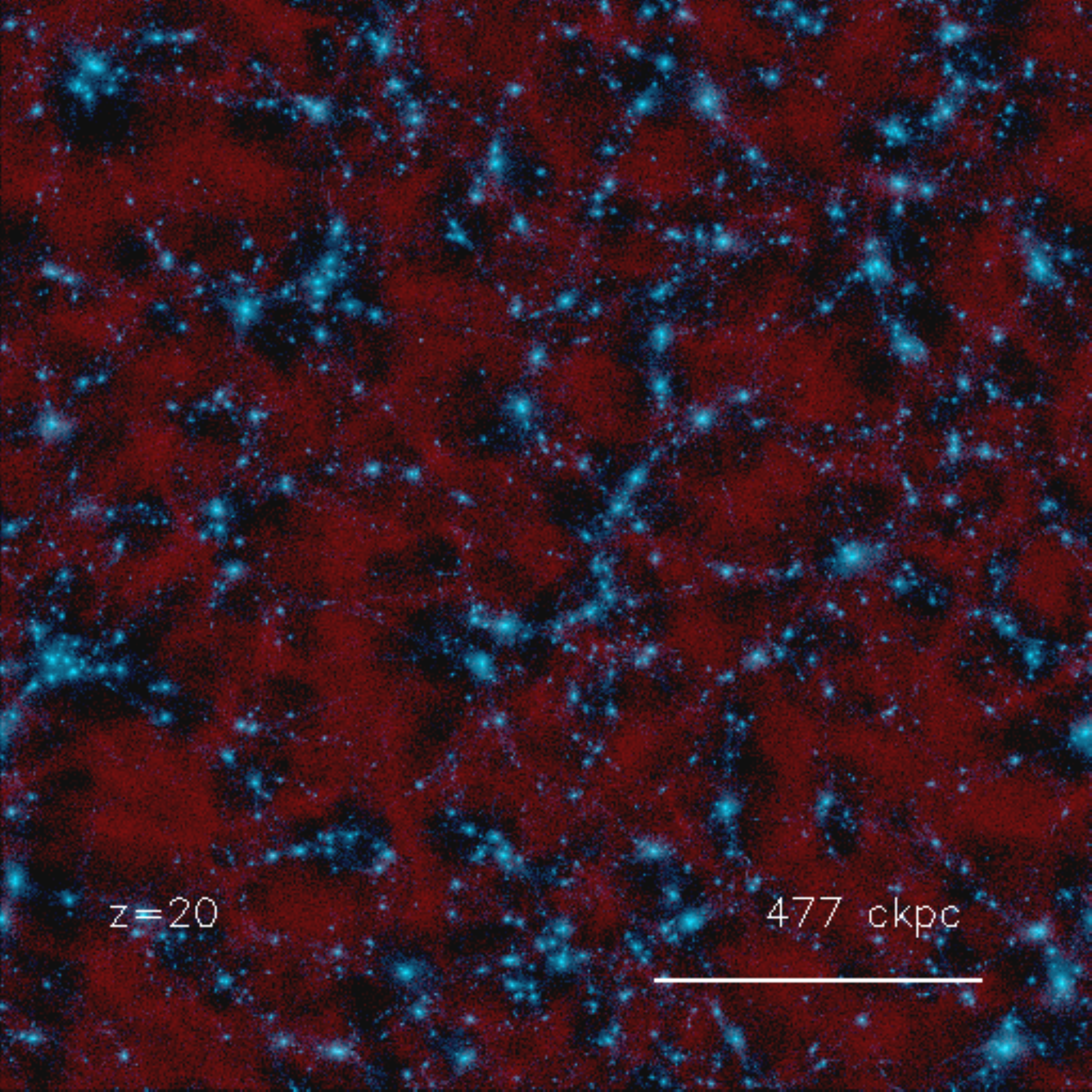}
 \end{center}
    \caption{Slice of a matter-antimatter simulation of a small volume in the D-M universe using a modified version of the  \texttt{RAMSES} simulation code. The thickness of the slice is 20\% of the simulation box, of comoving size 1 $h^{-1}$ cMpc. The condensed structures (represented in light blue) are all made of matter, while antimatter is spread in nearly homogeneous extended halos (represented in red) over half of the volume. The other half of the volume is occupied by depletion zones, surrounding the matter galaxies and clusters, isolating them from antimatter. These depletion zones are a consequence of the gravitational polarization expected between positive mass and ``negative mass" objects. We note that antimatter clouds percolate, \emph{i.e.} it is possible to roam at infinity without leaving the antimatter cloud, and the same property is respected for the depletion zones surrounding matter.} \label{fig:ramses_slice}
\end{figure}

In order to introduce the properties of rotation curves and virial velocity distributions in the D-M cosmology, let us first consider the following example of cluster configuration, represented on Fig.\,\ref{fig:ramses_slice}. This configuration was obtained in a 3D simulation using a modified version of the \texttt{RAMSES} code \citep{Teyssier_2002}, that we discuss more precisely in the next section. This figure represents a small cluster configuration in the D-M universe at redshift $z \approx 20$ for self-gravitating particles, at this stage without dissipation. The full tomography of the matter, antimatter, and matter+antimatter configurations can be found as supplementary material at the following link\footnotemark[2].
\footnotetext[2]{\url{https://youtu.be/rdIOCoy8QPM}}

An animation of the formation of structures centered on a massive cluster of the simulation can be found at the following link\footnotemark[3].
\footnotetext[3]{\url{https://youtu.be/aqyuDYrwyBQ}}

In Fig.\,\ref{fig:ramses_slice}, it can be seen that while matter has the usual clustering properties in planes, filaments and nodes (where globular clusters, galaxies and clusters of galaxies accumulate), and is concentrated in relatively small regions, antimatter has a completely different distribution, occupying nearly exactly half of the total volume, with a much more homogeneous density than matter. In addition, a new element is apparent: empty or very low-density depletion zones surround matter, isolating it from antimatter and occupying also about half of the total volume. We note that while it will be in most occasions possible to define rather well-defined isolated matter galaxies and clusters, this will not be the case for antimatter clouds, which are percolating. Strictly speaking, it is therefore incorrect to speak of ``domains" for antimatter, since there is path continuity at all distances within the antimatter clouds. The same property of percolation is respected by the depletion zones, which are therefore organized more in tubes surrounding matter structures than in spheres.

\begin{figure}[]
 \begin{center}
\includegraphics[width=0.9\linewidth]{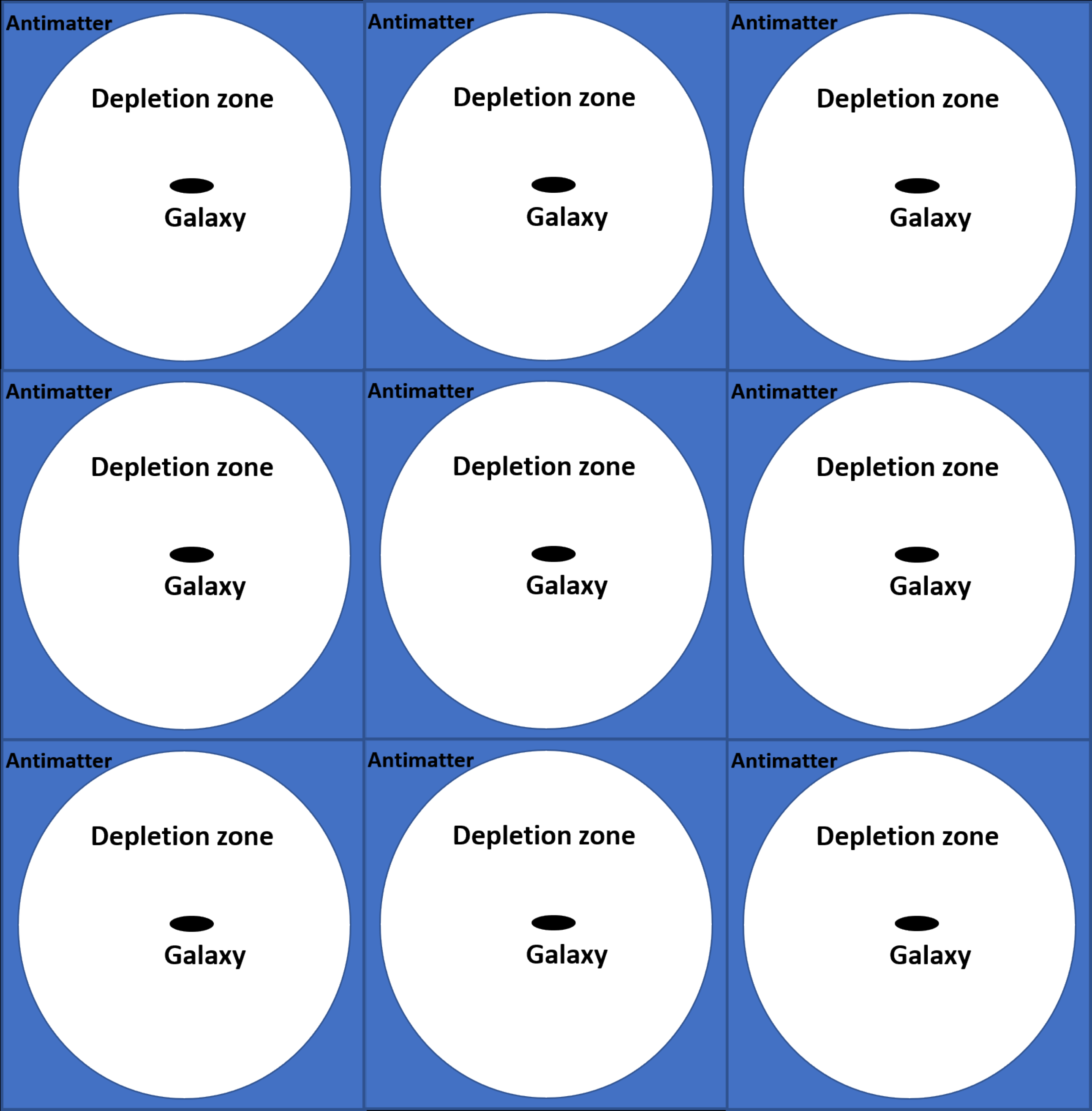}
 \end{center}
    \caption{Schematic representation of a periodic structure of identical galaxies and their surroundings in the D-M universe. A galaxy core is represented by a condensed object at the center of a cubic cell, while a depletion zone, occupying about 50 \% of the cell volume, surrounds the galaxy. Antimatter is spread out beyond this depletion zone, on the outskirts of the cell, occupying also nearly 50 \% of the cell volume, although its volume appears smaller in the 2D projection.} \label{fig:periodic_dm_galaxy}
\end{figure}

Let us now show that both these antimatter clouds and the depletion zones between matter and antimatter lead to an additional surface gravity compared to the Newtonian expectation. For this, we first use  a simple analytical model, represented on Fig. \ref{fig:periodic_dm_galaxy}, where the D-M universe is represented as the periodic repetition of elementary cubic cells, in which half the volume is occupied by a galaxy (located at the center of the box) and the surrounding almost spherical depletion zone, while the antimatter cloud occupies the other half of the volume in the outer part of the box. Although this is clearly an oversimplification, since in this representation all structures are supposed to have the same mass and size, it will enable us to evidence the salient features of galactic rotation curves in the D-M universe.


More precisely, we postulate that each galactic cell is constituted by a cubic box of volume equal to the average volume per galaxy. We represent the content of this box as the sum of three contributions:
\begin{enumerate}[label=\alph*)]
\item{A galaxy concentrated in a very small volume at the center of the box, the density of the galaxy being much higher than the average matter density (typically a factor 200 within the virial radius);}
\item{A depletion zone, mostly empty of matter and antimatter, around the galaxy.
In the galactic box, the volume of the depletion zone is about half of the total volume of the box when antimatter and matter in the box have equal and opposite ``masses", as supposed in D-M;}
\item{Surrounding this depletion zone, a region filled with antimatter with nearly constant density and a volume equal to half of the total volume of the box.}
\end{enumerate}
The whole space is then represented as a collection of such cubes in a periodic geometry.

We note that although the antimatter clouds occupy half of the volume of the box, the radius of the depletion zone extends almost to the confines of the box.
In order to show this, let us derive, in our approximation of spherical symmetry, the extent of the depletion zone delimiting the zones of matter (mostly condensed) and antimatter (spread out almost uniformly). The constraints are the following:
\begin{itemize}
\item{The total mass of the ``galactic cell" is zero, \emph{i.e.} $m_{+} = m_{-} \equiv m$}
\item{The depletion zone is spherical and of radius $r_d$}
\item{The linear size of the individual galactic cell is $L$, and its total volume is therefore $L^3$.}
\item{The antimatter, due to its internal repulsion, has a constant density $\rho_{-}$.}
\end{itemize}
The edge of the depletion zone is defined by the condition that the total gravitational force on an antimatter particle becomes zero. In order to express the repulsive force exerted by the antimatter cloud on an antiparticle at the edge of the depletion zone, we use the fact that the force created by the antimatter cloud of uniform density would be zero were it not for the spherical depletion zone. On the other hand, creating the depletion zone by removing an homogeneous sphere of antimatter creates a force directed toward the center of the box, and its intensity can be calculated by Newton's shell theorem stating that the situation is equivalent to that where the antimatter mass of the depletion volume is concentrated at the center. The spherical depletion zone will therefore create an attractive force as it if were concentrated at the origin and with a mass $\frac{4}{3} \pi \rho_{-} r_{d}^3 $. Since we want the total gravitational force on an antimatter particle at the edge of the depletion zone to be zero, it means that this mass must be equal to $m_{+}$.

On the other hand, the remaining mass of antimatter, outside the depletion zone, is:
\begin{equation}\label{mass_antimatter_cloud}
m_{-} = \rho_{-} ({ L}^3 - \frac{4}{3} \pi r_{d}^3)
\end{equation}
By our first condition, it must be equal to the positive mass $m_{+}$ of the galaxy.

Overall, we have therefore the conditions relating $\rho_{-}$ and $r_d$:
\begin{equation}\label{depletion_zone_radius}
\rho_{-} ({ L^3 - \frac{4}{3} \pi r_{d}^3) = \frac{4}{3} \pi \rho_{-} r_{d}^3 = m_+ = m_-}
\end{equation}
Therefore, in our approximation of spherical symmetry, the antimatter cloud occupies half the volume of a simulation cube, and the depletion zone extends until:

\begin{equation}\label{depletion_radius_2}
r_d = \sqrt[3]{\frac{3}{8\pi}} { L}  \approx 0.492 { L}
\end{equation}

This means that, counterintuitively, the depletion zone almost reaches the confines of the average individual box around a galaxy.
It is important to note that the situation that we have represented on Fig. \ref{fig:periodic_dm_galaxy} has not a spherical symmetry, notably concerning the antimatter cloud. Also, in the actual 3D equilibrium configuration, the depletion zone of a cell connects with that of the adjacent cells on its three axis.

\subsection{Analytical approximation of the rotation velocity}

Let us now estimate the rotation velocity for a matter test particle as a function of the distance to the galaxy in the preceding configuration.
The galaxy of mass $m$ obviously contributes, for a test body located at the distance $r$ from the galaxy center, to an acceleration equal to $Gm/r^2$, as in the Newtonian case.

 For a matter particle, the gravitational force due to the configuration of the antimatter cloud and the depletion zone surrounding the galactic core can be estimated as follows. We first consider the configuration where antimatter of ``negative mass" (see \citealt{Manfredi2018} for a precise definition), is uniformly distributed in the box. The gravitational force exerted by this uniform antimatter cloud is then zero everywhere, by symmetry.
Then, we represent the depletion zone as the sum of a homogeneous sphere of positive density (equal and opposite to that of antimatter) covering the volume of the depletion zone, which will compensate the uniform background of the negative mass fluid.

\begin{figure}[]
 \begin{center}
\includegraphics[width=\linewidth]{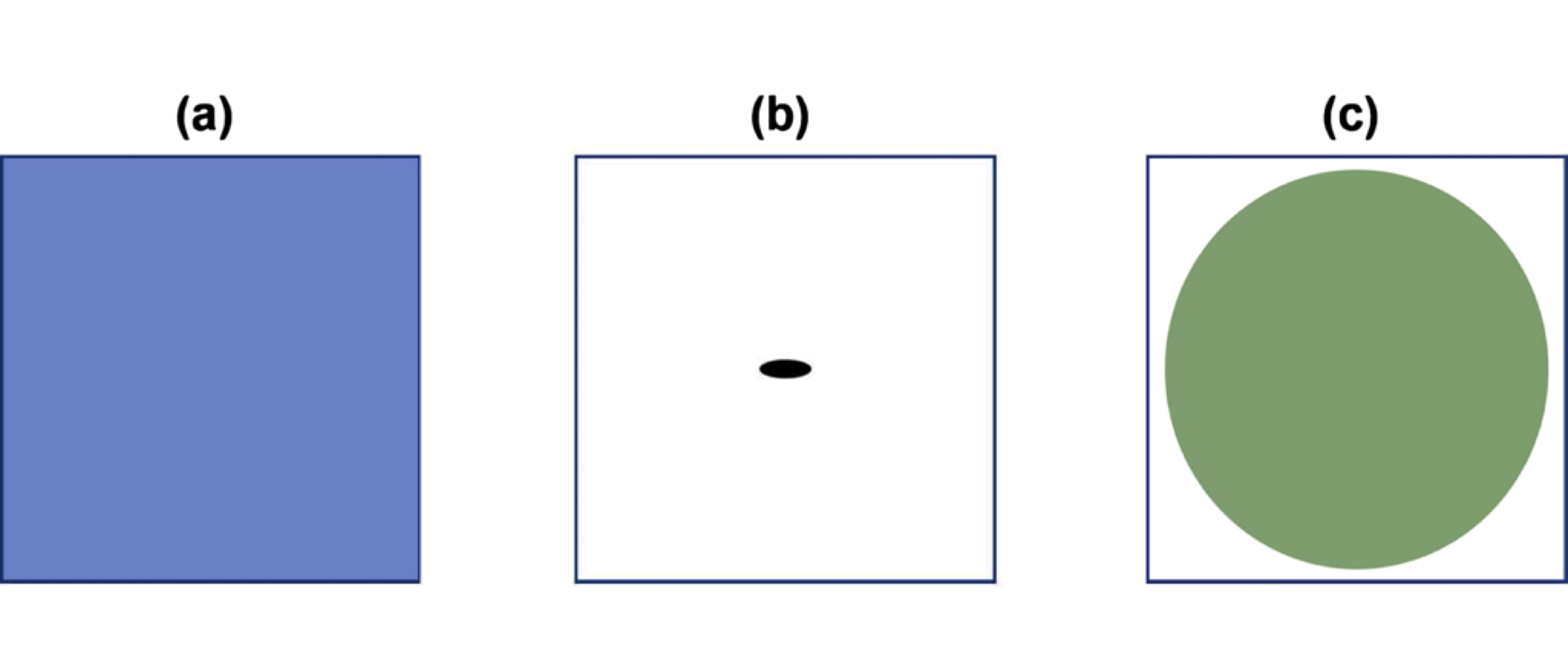}
 \end{center}
    \caption{Our approximation of a galaxy represented on Fig. \protect{\ref{fig:periodic_dm_galaxy}} can be represented
    as the sum of three cubes:
    a cube (a) with uniform ``negative mass" density of total mass $-2m$,
    a cube (b) with only the galaxy of mass $+m$ at the center,
    a cube (c) with a sphere occupying half the volume of the cube and of homogeneous positive density of matter, equal
    and opposite to that of the first cube, that together with (a) will create the empty depletion zone.} \label{fig:Sum_3_cubes}
\end{figure}

For each individual cubic cell, the situation shown in Fig.\,\ref{fig:periodic_dm_galaxy} can therefore be expressed as the superposition of three cubes (Fig.\,\ref{fig:Sum_3_cubes}):
\begin{enumerate}[label=\alph*)]
\item{A cube with ``negative mass" equal to $-2m$,
\emph{i.e.} {\it twice} the mass of the galaxy, and uniform density (zero gravitational fields $\vec{g}_{+}$ and $\vec{g}_{-}$ everywhere);}
\item{A cube with the mass $m$ of the galaxy at the centre, \emph{i.e.} half of the total mass of the antimatter in the first box.
This galactic point-like mass will produce the classical Newton force field $\vec{g}_{+} = -Gm\vec{r}/r^3$ (and $\vec{g}_{-} = -\vec{g}_{+}$);}
\item{A cube with the same positive mass $+m$ as the galaxy, but uniformly distributed over the spherical depletion zone. This positive mass will compensate the negative mass $-m$ of the first cube contained in the same sphere, thus creating the empty depletion zone, with a volume 50\% of the box.}
\end{enumerate}

As is well known, the homogeneous matter sphere in the third cube will generate a harmonic restoring force, which is superimposed on the Newtonian force of the galaxy, supposed to be a point galaxy. We note that the second and third cubes have a total mass of $2m$, compensating the ``negative mass" $-2m$ of antimatter in the first cube.
Space is supposed to be covered by adjacent such cubes (see Fig. \ref{fig:periodic_dm_galaxy}).

At a distance $r$ from the center of the box (and the center of the galaxy), the gravitational field created by this homogeneous sphere of positive density will then be:
\begin{equation}\label{g_homogeneous_sphere}
\vec{g}_{+} = -\frac{Gm \vec{r}} {r_d^3}
\end{equation}

where $m$ is the positive mass added to create the depletion zone and $\vec{r}$ is the radial distance vector to the galactic center.
The total gravitational field $\vec{g}_{+}$ exerted on a matter particle orbiting at a distance $r$ from the galaxy center is therefore:
\begin{equation}\label{total_g}
\vec{g}_{+}  = -\frac{G m\vec{r}}{r^3} -\frac{Gm \vec{r}}{r_d^3}
\end{equation}
while the relation between orbital velocity and this force is:
\begin{equation}\label{rotation_velocity_1}
\frac{v^2}{r}= \frac{G m}{r^2} +\frac{Gm r}{r_d^3}
\end{equation}
The orbital velocity as a function of the distance $r$ to the center of the galaxy can be simply deduced from the above expression, and reads as:
\begin{equation}\label{rotation_velocity_2}
v(r)=\sqrt{\frac{G m}{r} +\frac{Gm r^2}{r_d^3}}
\end{equation}

\begin{figure}[]
 \begin{center}
\includegraphics[width=0.9\linewidth]{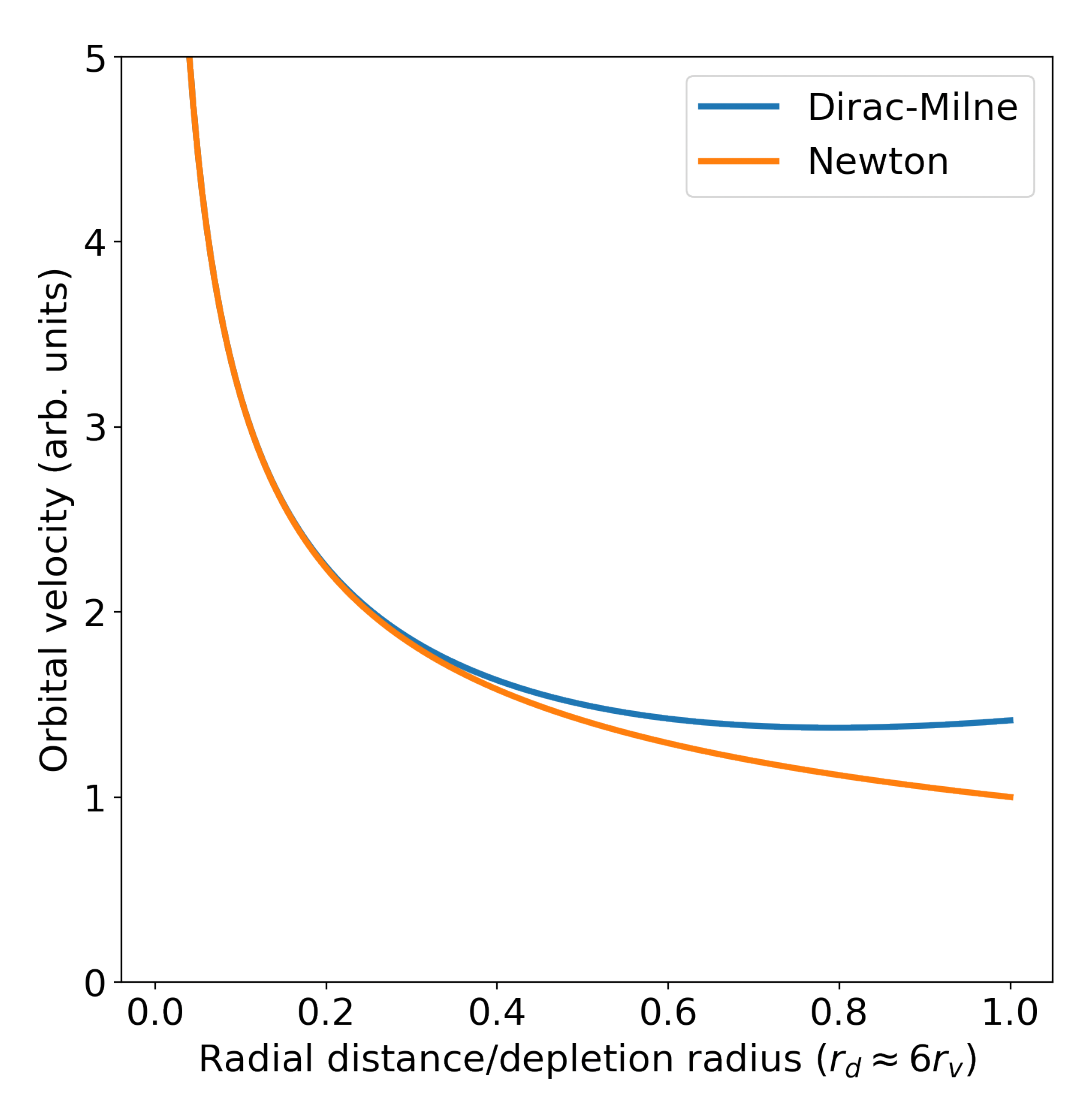}
 \end{center}
    \caption{Rotation velocity predicted in the D-M universe for a point mass located at $r=0$
    and surrounded by a depletion zone and the same mass of antimatter (with opposite ``sign").
    Instead of the expected Newtonian velocity (see Eq.\,\eqref{g_homogeneous_sphere}), decreasing to zero, D-M predicts that the
    rotation velocity for a condensed object of positive mass will be nearly constant (see Eq.\,\eqref{rotation_velocity_2})
     above the characteristic distance approximately equal to
    half the radius $r_d$ of the depletion zone, or $\approx 3$ virial radii.} \label{fig:dm_rotation_velocity}
\end{figure}

Remarkably, this function has a derivative that vanishes at  $r=2^{-1/3} r_d \approx 0.79 \, r_d$, and is almost constant for all $r \gtrsim r_d/2$.
This function is represented in Fig. \ref{fig:dm_rotation_velocity}, where we also plot the expected orbital velocity assuming 
there were no depletion zones nor antimatter clouds on the outskirts.
This figure shows that for such idealized galaxies, the D-M orbital velocity reaches a nearly constant plateau, even rising very slightly near the end of the depletion zone in our approximation of spherical symmetry. Note, however, that  for $r>r_d$, the spherical approximation breaks down as we have seen in Sec.\,\ref{sec:config_sphere}. Additionally, beyond the depletion zone, the antimatter cloud is present.

Since the depletion zones occupy about 50\% of the total volume and since the baryonic density in a D-M universe is defined by the baryon/photon ratio
$\eta \approx 8 \times 10^{-9}$ \citep{Sethi_1999, Benoitlevy}, an observer assuming that the \LCDM cosmology is valid will interpret the depletion zone as a halo with a density equal to the matter density of the D-M universe, \emph{i.e.}:
\begin{equation}\label{dirac_milne_density}
\frac{\rho_m}{\rho_c} = \frac{\eta_{\rm D-M}}{\eta_{\rm \Lambda CDM}} \times \Omega_{baryon}^{\rm\Lambda CDM} \approx \frac{8 \times 10^{-9}}{6 \times 10^{-10}} \times 0.049 \approx 0.65
\end{equation}
where $\eta = n_B/n_{\gamma}$ is the number of baryons per photon in the corresponding cosmology, and $\rho_c$ is the critical density for the \LCDM universe at the present epoch.

Now, the precise characterization of the virial radius depends on the cosmological model used and notably on the inflows between the redshift of initial collapse of a structure and the present epoch. The overdensity $\Delta_c$ at the virial radius for \LCDM is usually approximated as $\Delta_c \approx 100 \rho_c$ \citep{Shull_2014}, so that a nearly constant rotation velocity will be observed for distances:
\begin{equation}\label{radius_flat_rotation}
r \gtrsim \frac{r_d}{2} \approx \frac{1}{2}\sqrt[3] {\frac{\Delta_c}{\rho_m}}\,  r_v  \approx \frac{1}{2}\sqrt[3] {150} \, r_v \approx 2.6\, r_v
\end{equation}
where $r_v$ is the virial radius. Therefore, the flat rotation curves, instead of being due to an invisible halo of slowly decreasing density,
are in fact due in D-M to the asymmetric configuration of matter and antimatter that we have described above.

Indeed, the D-M universe tells us that in order to derive the galactic rotation
curves and the virial velocities in a cluster of galaxies, it is necessary to take into account not only the matter present
between us and the center of the galaxy or the cluster, \emph{i.e.} matter ``below our feet", but also (anti)matter ``above our heads",
which may be interpreted as an External Field Effect (EFE) \citep{SEP_EFE_2020}.
This is indeed very counterintuitive, and may even be considered in contradiction with the Newtonian shell theorem\,\citep{Newton_1760},
stating that inside a spherical shell, spacetime must be Minkowskian, \emph{i.e.} the gravitational field must be zero.
We further discuss  this apparent contradiction in Appendix\,\ref{app:A}.

The approximation of spherical symmetry allowed us to show that flat rotation curves are expected within a large fraction of the volume of the depletion zone. Using a numerical simulation, we now explore in more precision the edge effects created by the antimatter clouds, which are in fact asymmetric.

\subsection{Numerical study of the depletion zone}
\label{sec:config_sphere}
\begin{figure}[]
\centering
\includegraphics[width=0.9\linewidth]{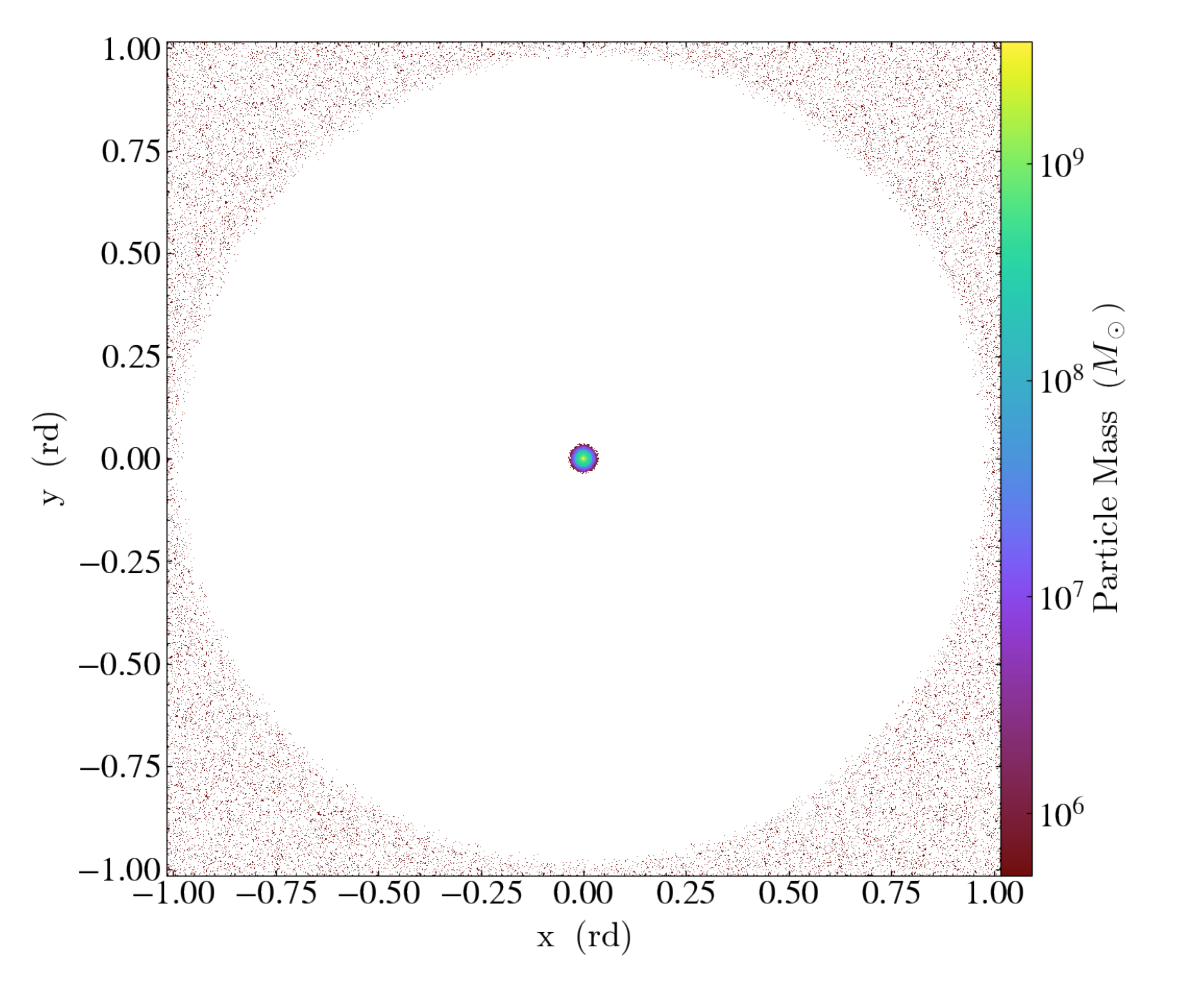} 
\caption{Configuration studied in this simulation. A matter galaxy, located at the center of the box, is surrounded by a depletion zone,
extending over half the volume of the box, and by an antimatter cloud (represented in magenta), extending to the outskirts of the box over the other half
of the volume.} \label{fig:config_sphere}
\end{figure}

In this subsection, we use our  modified version of \texttt{RAMSES} (see Section~\ref{sec:simulations} for details) to study numerically the spherical approximation of the previous subsection. We consider a galaxy shell configuration following Section \ref{sec:repart}, as described in Fig.\,\ref{fig:config_sphere}.

More precisely, we considered an initial configuration where a central galaxy core, confined in a radius equal to $\approx 5\%$ of the simulation box length, presents an isothermal density and velocity profile. The same mass of antimatter is uniformly distributed  at $r>r_d \approx 0.5 L$, (see Eq.\,\eqref{depletion_radius_2}), but its initial velocity is zero. The boundary conditions for the simulation box are periodic, which means that the actual configuration simulated is that of Fig.\,\ref{fig:periodic_dm_galaxy}, repeated out to infinity.

\begin{figure}[]
\centering
\includegraphics[width=0.9\linewidth]{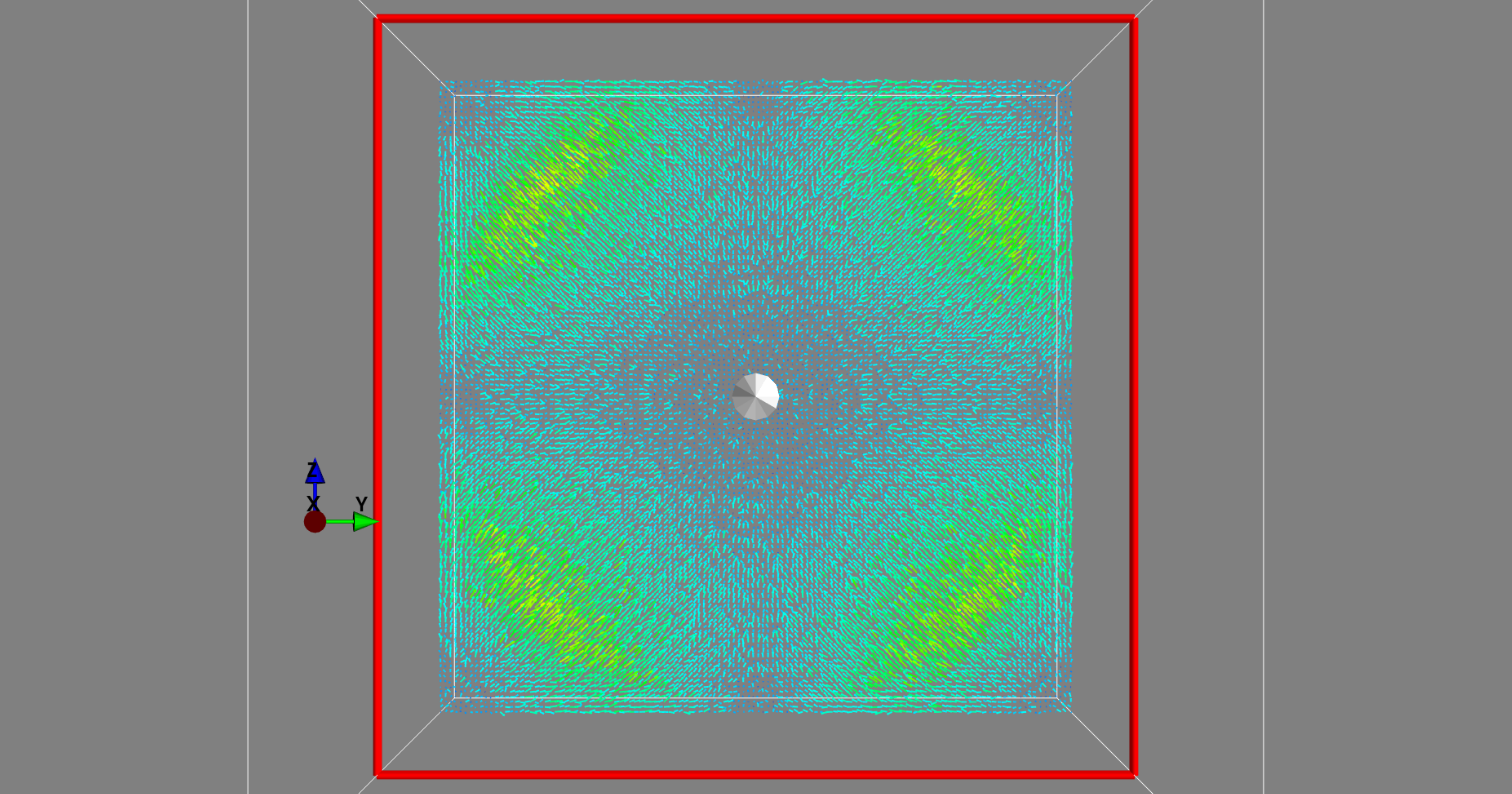} 
\includegraphics[width=0.9\linewidth]{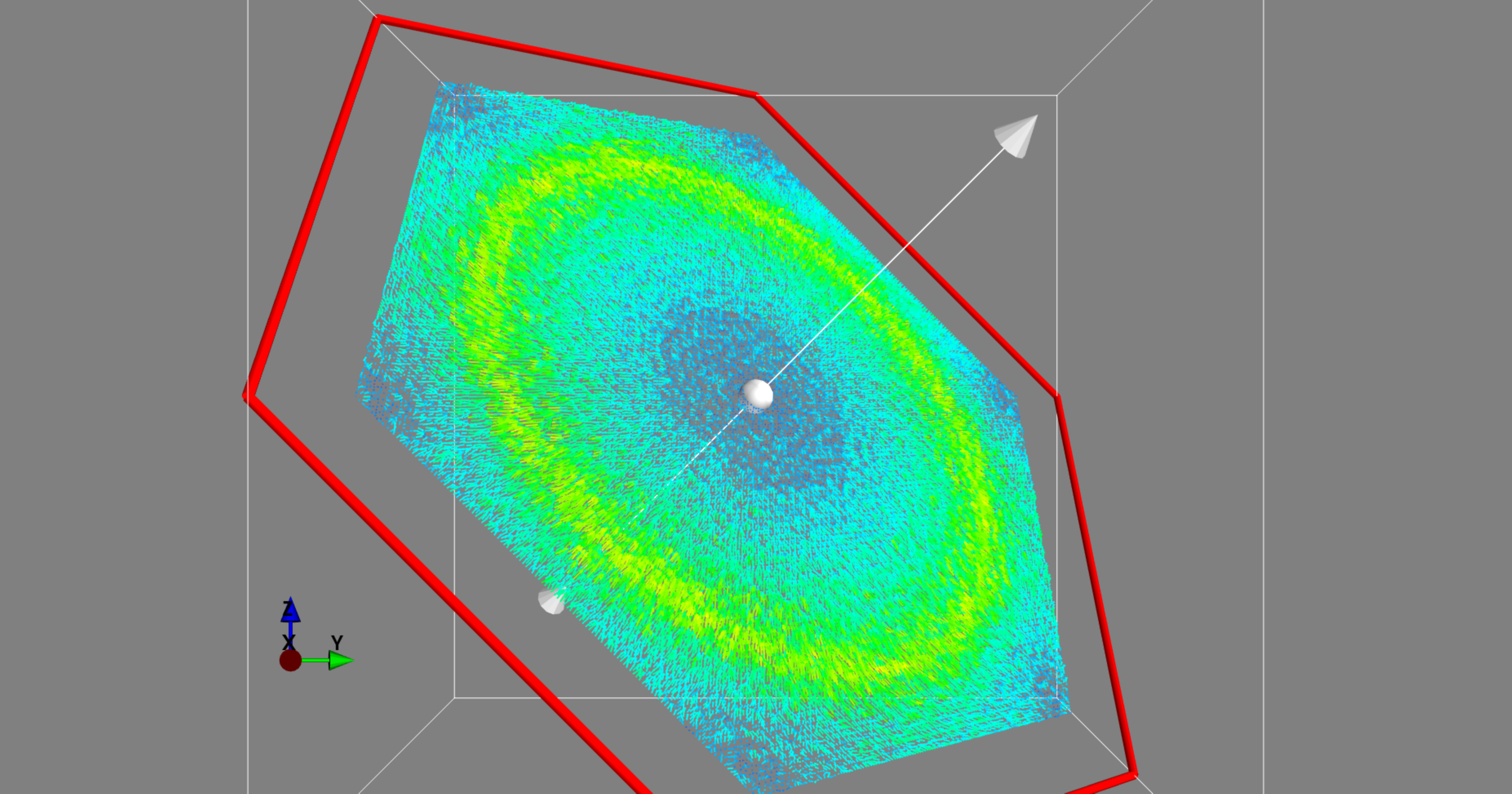}
\caption{Gravitational field $\vec{g}_{\text{am}}$ created by antimatter on two planes defined by their normal vector (1,0,0) for the top panel and $\frac {(1, 1, 1)} {\sqrt{3}}$ for the bottom panel. The color scaling indicates the strength of the field. The confining gravitational field created by the antimatter cloud is more symmetric and has a higher intensity in the bottom panel, which represents the field in one of the diagonal planes of the cube.} \label{fig:antimatter_in_config_sphere}
\end{figure}

Using \texttt{RAMSES} on a $128^3$ grid, we checked that this initial configuration does not evolve noticeably. The gravitational field $\vec{g}_{\text{am}}$ created by antimatter is represented in Fig.\,\ref{fig:antimatter_in_config_sphere}. Due to the non-spherical distribution, the antimatter field is seen to be stronger on the diagonals of the cube for a plane with normal parallel to one of the axes of the cube, \emph{e.g.} with a normal vector (1, 0, 0). In the bottom panel of the same figure, we have represented the antimatter field for a plane along one of the diagonals of the cube, \emph{e.g.} with a normal vector $\frac {(1, 1, 1)} {\sqrt{3}}$. For such planes, the confining field created by antimatter is more symmetrical, in a ring configuration, and stronger than the field present in the plane parallel to one of the faces of the cube.

\begin{figure}[]
\centering
\includegraphics[width=\linewidth]{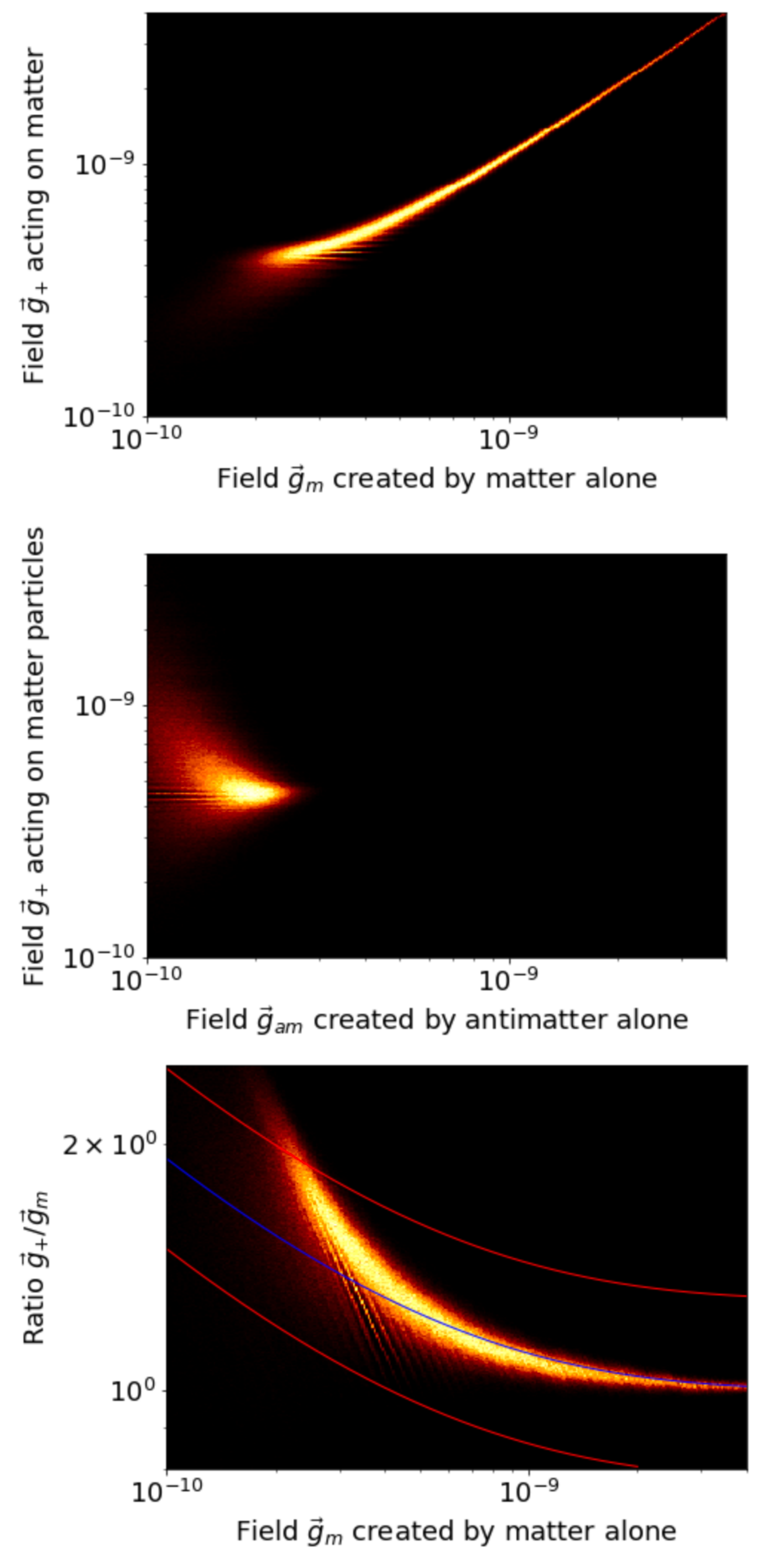}
\caption{Top panel: Total gravitational field acting on a matter particle as a function of the gravitational field created by matter alone. A Newtonian regime can be observed for the high field region of the figure, with a gradual transition from the Newtonian regime to a MOND-like regime at an acceleration of $\approx 1.0 \times 10^{-9}$ m/s$^2$.
Middle panel: Total gravitational field acting on a matter particle as a function of the gravitational field created by antimatter alone. The antimatter field is rather peaked, with an average value of $\approx 2 \times 10^{-10}$ m/s$^2$.
Bottom panel: Ratio between the total gravitational field acting on a matter particle and the Newtonian field (created by matter only), as a function of the gravitational field created by matter only. Almost exactly unity in the Newtonian regime (at high values of the field), this ratio gradually increases and reaches a factor of $\approx 2$ at the end of the depletion zone. The MOND interpolating function used in \cite{Lelli_BTFR_2019} for a value of the $a_0$ parameter 1.85 $\times 10^{-10}$ m/s$^2$ has been superimposed on the simulation data, while the two adjacent curves represent the 1-$\sigma$ error of 0.11 dex found by these authors.} \label{fig:Sphere_grav_field}
\end{figure}

In order to quantify this statement, we represent in Fig.\,\ref{fig:Sphere_grav_field} the total gravitational field experienced by a matter particle, created by both matter and antimatter, as a function of the gravitational field created by matter alone. These gravitational fields are calculated on a $128^3$ Cartesian grid  (about 2 million cells), and the quantities estimated on the matter region, as defined by the distance to the galactic center, which is the only cut applied for the selection. 
A Newtonian regime can be observed in the high field region of the figure, with a gradual transition from the Newtonian regime to a MOND-like regime beginning at an acceleration of $\approx 1.0 \times 10^{-9}$ m/s$^2$. Fitting an $a_0$ parameter according to MOND interpolating formulas listed in \cite{Famaey_McGaugh_2012} results in values of the $a_0$ parameter ranging between $\approx 1.5$ and $\approx 5 \times 10^{-10}$ m/s$^2$, depending on the parametrization chosen.
Using as an example the parametrization of \cite{Lelli_RAR_2017}, we present in the bottom panel of Fig.\,\ref{fig:Sphere_grav_field} the ratio of the gravitational field acting on a matter particle, $g_{+}$, to the gravitational field created by matter alone, {\it i.e.} the Newtonian expectation, $g_m$. Superimposed on
this 2D distribution, we have plotted the analytical expression of \cite{Lelli_RAR_2017} for values of the $a_0$ parameter between 1.5 and 3.0 $\times 10^{-10}$ m/s$^2$. We have also indicated the upper and lower limits of 0.11 dex (1-$\sigma$ error) reported by \cite{Lelli_RAR_2017} in their analysis.

It should be noted, however, that the proximity with the value of the MOND parameter equal to $\approx 1.2 \times 10^{-10}$ m/s$^2$ is coincidental, as other values of the mass of the central structure in our simulation would have changed the value of the fitted parameter $a_0$.
We further discuss this crucial aspect in Section\,\ref{sec:MOND}, which is dedicated to the simulations of ``clusters'' of mass between $2 \times 10^{10} M_{\sun}$ and $1 \times 10^{16} M_{\sun}$, and where we study the evolution and value of the $a_0$ parameter predicted by D-M as a function of redshift and total mass.

%


\subsection{Self-consistent model of the depletion zone}

In the present section, in order to confirm the qualitative and numerical analysis of the depletion zone presented in the two preceding subsections, we extend the analysis of \cite{Manfredi2020} to incorporate the notion of depletion zone and inhomogeneity of the antimatter cloud. Using the Poisson equations (\ref{eq:poiss_diracmilne1}) and (\ref{eq:poiss_diracmilne2}), we consider a spherically symmetric geometry, where all quantities depend only on the radius $r$ and the Laplacian operator is defined as $\nabla^2 \phi= {1\over r^2} \partial_r (r^2 \partial_r \phi)$, where $\partial_r$ is a radial derivative. We want to solve these Poisson equations for a typical situation where a positive high-density mass (``galaxy") is located in a  small spatial region located near $r=0$ and is described by the distribution $\rho_{+}(r)$.
In contrast, negative masses are supposed to be thermalized (at low temperature $T$) and described by a Boltzmann distribution:
\be
\rho_{-}(r) = \rho_0 \exp\left(\frac{-m\phi_{-}+\mu}{k_B T}\right) ,
\label{eq:rhominus}
\ee
where $k_B$ is Boltzmann's constant, $\rho_0$ is a reference density, and $\mu$ is a chemical potential that will be chosen so that positive and negative masses are present in equal amounts in the computational box, \emph{i.e.}: $4\pi \int \rho_{+} r^2 dr =4\pi \int \rho_{-} r^2 dr $. We note that the mass $m$ appearing in Eq. (\ref{eq:rhominus}) is the {\em passive} gravitational mass, which in the D-M model is always positive \citep{Manfredi2018}.
Then, Eq.\,(\ref{eq:poiss_diracmilne2}) becomes a nonlinear Poisson equation for $\phi_{-}(r)$, which can be solved self-consistently, and the result substituted into Eq.(\ref{eq:poiss_diracmilne1}) in order to obtain $\phi_{+}(r)$.

However, as discussed at the end of Section \ref{sec:DM_rotations_curves}, the external field effects require a careful treatment of the boundary conditions. Indeed, as they are written, the Poisson equations (\ref{eq:poiss_diracmilne1})-(\ref{eq:poiss_diracmilne2}) describe an isolated system in otherwise empty space. But this is not the case for a cosmological setting for which the positive masses are localized in small regions, and the negative masses spread out almost uniformly across about 50\% of the available space.

We therefore solve Eqs.\,(\ref{eq:poiss_diracmilne1})-(\ref{eq:poiss_diracmilne2}) in a spherical region between $r=0$ and $r=R$, using the following boundary conditions:
\begin{eqnarray}
\phi'_{+}(R) = 2\, \frac{4\pi G \int_0^R \rho_{+}\, r^2 dr}{R^2}, \phi'_{-}(R) = 0 \
\label{eq:bc-phiplus}
\end{eqnarray}
where the apex stands for differentiation with respect to $r$.

For comparison, the Newtonian case is computed by solving: $\nabla^2\phi_{\rm Newt} = 4\pi G \rho_{+}$, with boundary condition:
$\phi_{\rm Newt}(R) = 0 $.

\begin{figure}[]
{\includegraphics[width=0.9\linewidth]{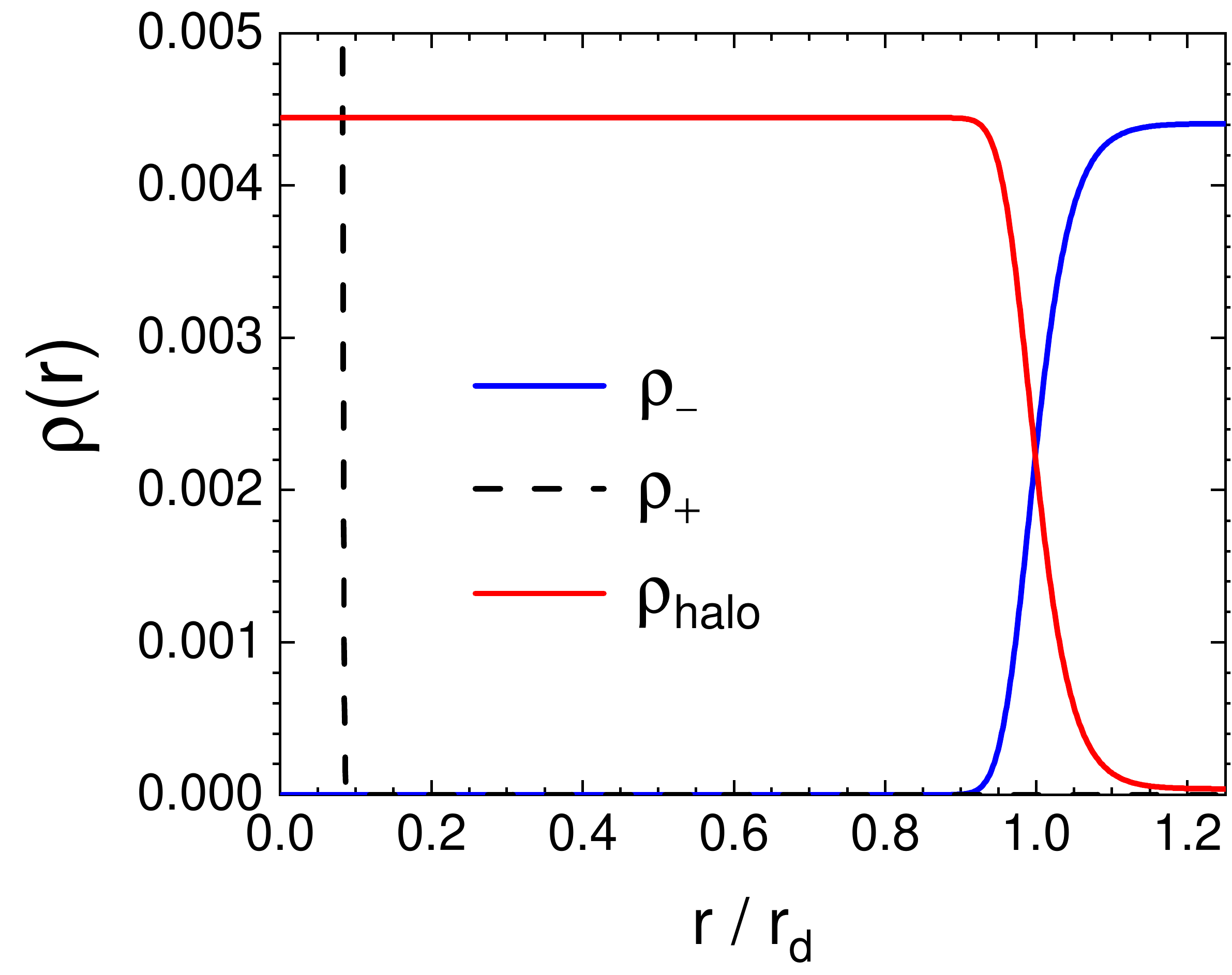}}
\caption{Matter density $\rho_{+}(r)$ (black dashed line), antimatter density $\rho_{-}(r)$ (blue line), and pseudo-dark-matter halo density:  $\rho_{\rm halo} = 2\bar{\rho}-\rho_{-}$ (red line).}
\label{fig:densities}
\end{figure}

\begin{figure}[]
{\includegraphics[width=0.85\linewidth]{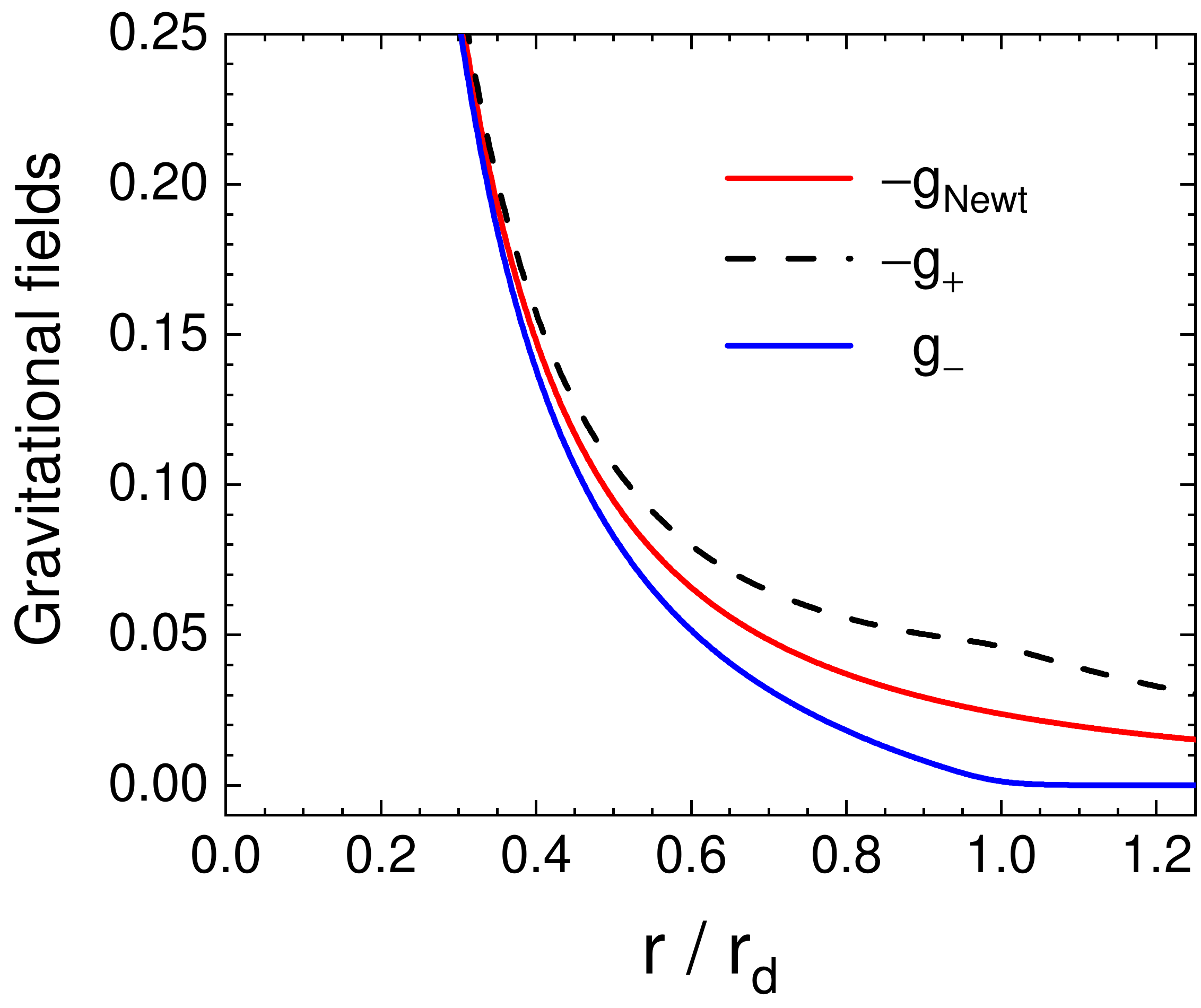}}
\caption{Gravitational  fields: $g_{-} $ (blue line), $-g_{+}$ (black dashed line), and $-g_{\rm Newt}$ (red line) .}
\label{fig:fields}
\end{figure}

\begin{figure}[]
{\includegraphics[width=0.85\linewidth]{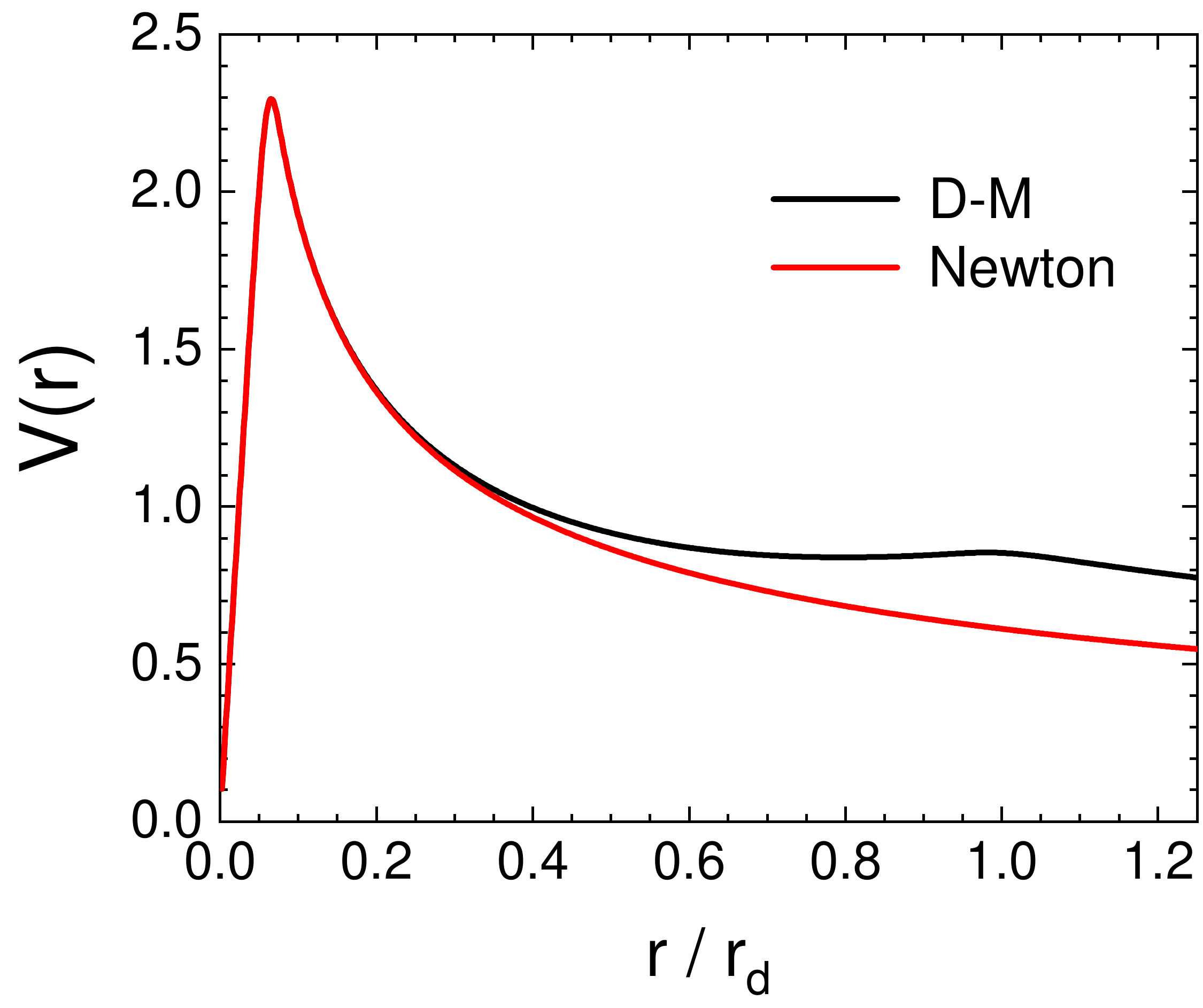}}
\caption{Rotation speeds for the D-M model (black line) and for standard Newtonian gravity (red line).
The decrease of the rotation velocity beyond $r/r_d =1$ is non physical, as the test particle would enter the antimatter cloud.}
\label{fig:speeds}
\end{figure}

In this subsection, we use units in which $4\pi G = 1$, and take $R=20$ and temperature $T=0.001$.
The depletion zone is clearly seen in the simulation results shown in Fig. \ref{fig:densities}, and extends up to $r_d \approx 15.8$. Thus, the ratio of the total volume to the volume of the depletion zone is approximately $R^3/r_d^3 = 2.04 \approx 2$, as expected.
On the outer border of the depletion zone, the gravitational field (defined as: $g = -\partial_r \phi$)  for matter is twice the Newtonian value, while the antimatter field vanishes within the entire depletion zone (see Fig. \ref{fig:fields}), justifying our hypothesis that the antimatter cloud is both very cold and homogeneous.
The rotation speeds are defined as: $v(r) = \sqrt{r\, | \phi'(r) | }$.
For the D-M model, the rotation curve flattens between approximately $r_d/2$ and $r_d$, also in accordance with our model (see Fig. \ref{fig:speeds}).

Finally, we point out that the Poisson equations (\ref{eq:poiss_diracmilne1})-(\ref{eq:poiss_diracmilne2}), with the boundary condition (\ref{eq:bc-phiplus}), are equivalent to adding a constant density $2\bar{\rho}$ on the right-hand side of both equations (where  $\bar{\rho} \equiv \langle \rho_{+} \rangle = \langle \rho_{-} \rangle$ is the average matter or antimatter density) and using Dirichlet boundary conditions $\phi_{\pm}(R) = 0$, \emph{i.e.} fixing the value of the potentials on the sphere ($r=R$). The conditions on the gradients (the fields), \emph{i.e.} Eq. (\ref{eq:bc-phiplus}), will be automatically satisfied because of Gauss's theorem.
Hence, the potential  $\phi_{+}$ acting on matter results from two sources:  the central ``galaxy" $\rho_{+}$ and the additional distribution $\rho_{\rm halo} \equiv 2\bar{\rho}-\rho_{-}$. The latter is analogous to the dark matter halo postulated in the standard CDM theory. This halo distribution is also represented in Fig. \ref{fig:densities}.

\section{Simulation results}\label{sec:simulations}

\paragraph*{} In order to validate the analytical approximations studied in Section\,\ref{sec:diracmilne}, we modified the Adaptive Mesh Refinement code \texttt{RAMSES} \citep{Teyssier_2002} in order to implement the gravitational behavior of matter and antimatter present in D-M, following Eqs.~\eqref{eq:poiss_diracmilne1}-\eqref{eq:poiss_diracmilne2}. We note that in most cosmological simulations, the average density is first calculated and subtracted from the local density in order to calculate the evolution of the scale factor $a(t)$, so that the Poisson equation is usually written $\nabla^2 \phi=4 \pi G a^2 \delta \rho$. In D-M, this is directly the case, since $\bar{\rho}=0$ in this matter/antimatter symmetric universe, so that $\delta \rho \equiv \rho(t,\boldsymbol{x})- \overline{\rho}(t)=\rho(t,\boldsymbol{x})$. The new element introduced in this D-M cosmological version of \texttt{RAMSES} is an extra set of particles with negative mass, which in turn introduces the new aspect of a depletion zone.

The acceleration of these negative mass particles is given by the gravitational potential of equation~(\ref{eq:poiss_diracmilne2}) (respectively, equation~(\ref{eq:poiss_diracmilne1}) for positive mass particles). The mass density of both particle species are projected separately onto the mesh with a cloud-in-cell interpolation, so that the mesh contains both the positive and negative mass density of the corresponding positive and negative mass particles. Once these two separate mass densities on the mesh are obtained, the positive and negative gravitational potentials are derived using the standard conjugate gradient algorithm of \texttt{RAMSES}.
The plus and minus accelerations are computed with a simple finite difference and attributed to the corresponding set of positive and negative mass particles with the cloud-in-cell interpolation of the accelerations computed on the mesh. All particles share a common level-based timestep obtained from the smallest Courant condition from particle velocities and free fall time. The cosmological time is a linear function of the scale factor $a(t)$, as in the Milne geometry\,\citep{Milne}.

\paragraph*{Numerical Setup}
The first simulation that we present involves a small simulation volume, of dimensions 1 $h^{-1}$ cMpc, discretized on a grid of $256^3$ cells. We allowed for up to two refinement levels, therefore leading to an effective resolution of
$\Delta x \approx 1 h^{-1} \rm ckpc$. According to \cite{Benoitlevy, Manfredi2020}, the initial size of matter and antimatter domains is of the order of 100 pc at $z \approx 1080$ ({\it i.e.} 100 ckpc at the present epoch).
Thus the resolution will capture these domains and their contrast at the initial times.  We ran the simulation from $z= 1080$ to $z = 0$. At a redshift of $z = 20$, 7769 halos were identified using the  AdaptaHOP algorithm\,\citep{Aubert_et_al_2004}, the development of structures in D-M occurring mostly before $z = 3$. This allows to sample structure characteristics such as size and velocity dispersion over a range of about three orders of magnitude in mass ($10^5 - 2 \times 10^8$ M$_\odot$). The total mass of matter present in this simulation is $\approx 3.3 \times 10^{10} M_{ \odot}$ (and the same mass of antimatter), which gives an individual equal mass for each particle of $\approx 3.9 \times 10^3 M_{ \odot}$.
For the initial conditions, we used a Gaussian velocity distribution with dispersion: $v_I=\sqrt{\frac{Gm}{r}} \approx 2 \text{ km/s}= \mathcal{O}(1)$ km/s.

We checked that, even when the particles are initially distributed on a uniform grid with $v=0$, the qualitative behavior of our results is not significantly modified. This is due to the fact that the initial contrast in density in D-M is already of order unity immediately after the CMB transition, leading to a very efficient virialization of the first structures within typically one Hubble time at that epoch, \emph{i.e.} a few million years.

In the simulation, we use the value $H_0=70$ km/s/Mpc for the Hubble parameter at the present time. We note that, unlike standard large-scale structure simulations, we did not impose a given initial power spectrum but rather a specific matter/antimatter pattern. In this first simulation, the matter-antimatter pattern for the initial condition was generated using an Ising code (with two states per spin, used here to represent matter and antimatter), on a $256^3$ Cartesian grid evolved for a few time steps starting from a random distribution, and using a temperature well below the second order transition. This procedure was used in order to create a fine-grained distribution where matter and antimatter both percolate in analogy with an emulsion. The characteristic size of the emulsion ($\approx 100$ pc at $z = 1080$) results from the annihilation and polarization of the matter and antimatter emulsion at much higher temperatures\,\citep{Benoitlevy}, and we do not describe this phase of the evolution here. This aspect will be treated elsewhere.
\paragraph*{Results} Analogous to observations of electrons and holes in semiconductors \citep{Tsidi1975}, this small-scale simulation exhibits a nearly empty depletion zone between the condensed clumps of matter and the extended clouds of antimatter.  Figure \ref{fig:ramses_slice} illustrates this geometric distribution, asymmetric between matter and antimatter. We also refer the reader to the supplementary movie available at the following weblink\footnotemark[4].
\footnotetext[4]{\url{https://youtu.be/aqyuDYrwyBQ}}
showing that unlike in \LCDM, structure formation in D-M starts very shortly after the CMB transition, as was already predicted by our earlier 1D simulations\,\citep{Manfredi2018, Manfredi2020}.
The configuration represented in Fig.\,\ref{fig:ramses_slice} is due to the fact that matter and antimatter repel each other, leading to a zone where almost no particles or antiparticles are present. As antimatter particles repel each other, antimatter spreads as much as it can without going close to matter, by which it is also repelled. Matter clusters and forms halos as in the conventional gravitational scenario. Indeed, by looking only at the matter zones, it is difficult at first glance to distinguish, with its planes, filaments and nodes, the matter configuration in D-M compared to \LCDM. Obviously, this qualitative statement deserves a more detailed study.

\begin{figure*}[]
 \begin{center}
\includegraphics[width=17cm]{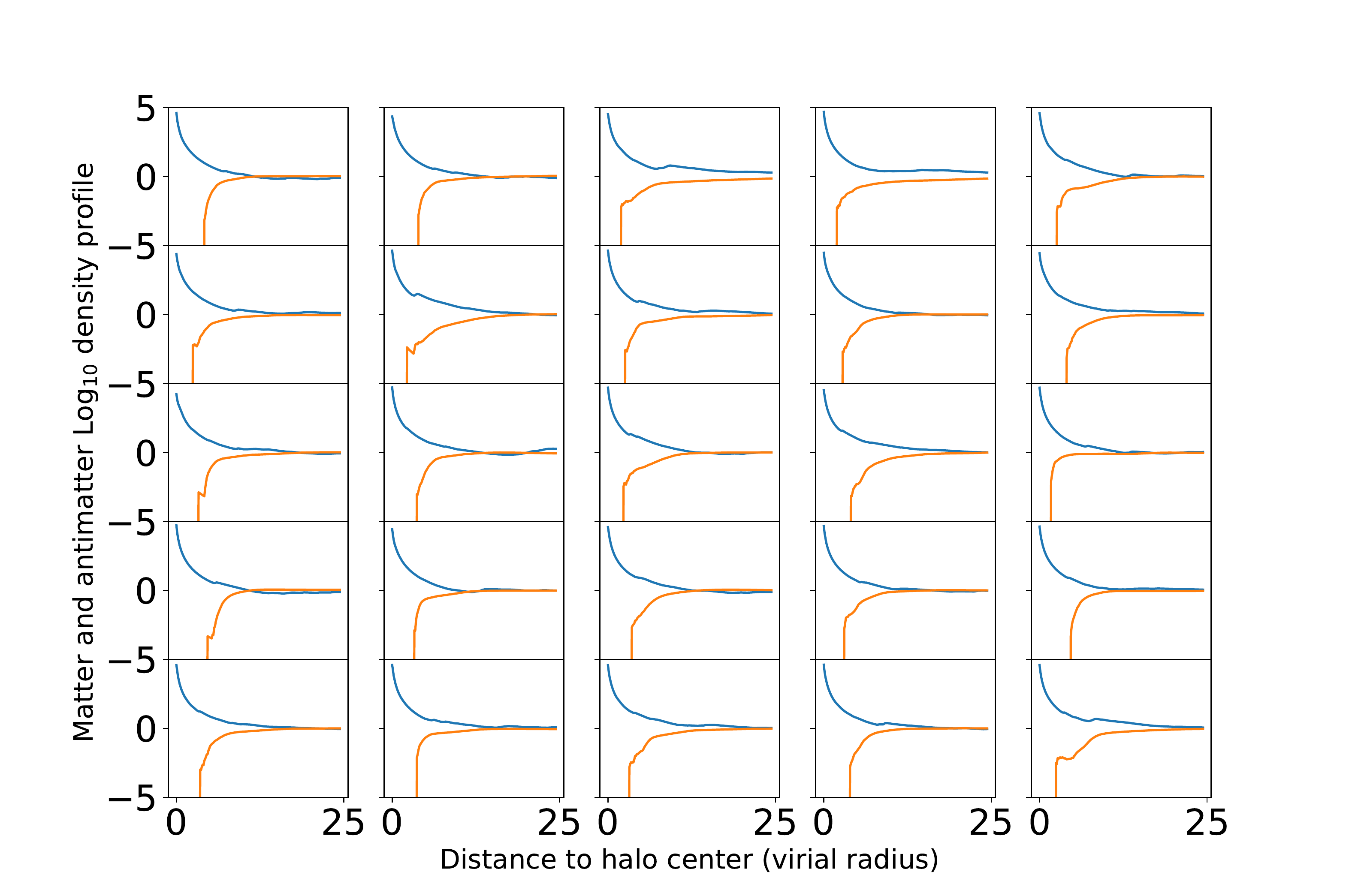}
 \end{center}
    \caption{In the present figure, we represent the logarithmic spherical density as a function of radius for both matter and antimatter for a set of 25 halos with mass $\approx 10^7$ M$_\odot$ in our \texttt{RAMSES} simulation. While the density of matter, represented in blue, is steeply decreasing from the halo center, the density of antimatter, represented in orange, is zero in this region. The logarithmic density is scaled to the average matter and antimatter density, equal in D-M at large scales, and converge (equal average densities for matter and antimatter) for almost all halos. It can be seen that the antimatter density becomes non zero after a few virial radii, and that the mass of antimatter becomes of the order of the halo matter mass after typically twice the radius at which antimatter density becomes non-zero. The distribution for 25 consecutive halos has been represented in the figure in order to show that this behavior is quite generic in D-M, although some variations can be seen in the distribution of antimatter around the matter halos.}
    \label{fig:Matter_antimatter_Halo_density_profile}
\end{figure*}

On the other hand, we can look at the common features of matter halos present in the 3D simulation. In Fig.\,\ref{fig:Matter_antimatter_Halo_density_profile}, we show the logarithmic density profile of matter and antimatter in 25 halos, with individual masses of the order of $10^7 M_\odot$ in our simulation. These density profiles share common features such as a bulge region where the matter density is steeply decreasing with the distance from the center of the halo, while the density of antimatter in the same region is zero. After a few virial radii, the antimatter density becomes non zero and the enclosed mass of antimatter rapidly increases and becomes of the same order as the mass of matter present in the halo, with nearly equal average densities. These extended antimatter clouds, present beyond the depletion zone, create an approximately harmonic restoring gravitational field, which adds its contribution to that of matter present in the halo, and lead to nearly flat rotation curves. This behavior is generic to most halos present in our simulation, although some isolated halos can present a somewhat different behavior when their vicinity includes more massive structures.
Depending on their antimatter environment, they might then appear as being completely devoid of dark matter when their close environment does not contain antimatter, with accelerations described by the expected Newtonian behavior or, on the contrary, as being submitted to strongly confining fields, created by the configuration of surrounding matter and antimatter clouds, as in the bottom panel of Fig.\,\ref{fig:antimatter_in_config_sphere}, and appear as almost entirely composed of dark matter.

\paragraph*{} As discussed in \cite{Manfredi2020}, the highly non-linear structure formation in D-M appears to give at the present epoch the same order of magnitude for the matter power spectrum at the peak scale as its \LCDM counterpart. However, at higher redshifts, the situation is quite different and bound structures appear much earlier in D-M than in \LCDM. This may indeed alleviate the constraints met by \LCDM in this regard\,\citep{Kovacs_ISW_2019, Asencio_ElGordo_2021, Lelli_Science_2021}.
It should be noted that in the evolution for redshifts between $z = 100$ and $z = 20$, the formation of structures in the upper range in mass may be underestimated as the size of the simulation box is very limited. Note also that in the present simulations, particles are dissipationless (behaving as dark matter particles), which may lead to a modification of the power spectrum at small scales.

A few additional remarks are in order: the average velocity of antimatter clouds is much smaller (typically by one order of magnitude, depending on redshift) than that of matter. In addition, these peculiar velocities for antimatter correspond to nearly global flows of cold antihydrogen (and antihelium) clouds, \emph{i.e.} the actual temperature of the antimatter clouds is much colder than that of matter. Also, as mentioned previously, antimatter clouds percolate with one another, which means that it is possible to travel continuously at infinity without leaving the antimatter region. Similarly, the depletion zones also exhibit percolation, meaning that matter structures are confined in tubes rather than in depletion spheres.

In order to better apprehend the gravitational influence of these extended antimatter clouds, we now turn to the study of the Tully-Fisher and the Faber-Jackson relations.


\section{Tully-Fisher and  Faber-Jackson relations in the D-M universe}\label{sec:TFR}
The Tully-Fisher relation \citep[TFR;][]{Tully_Fisher_1977} and Faber-Jackson relation \citep[FJR;][]{Faber_Jackson_1976} are phenomenological  scaling relations between the velocities inside galaxies and clusters and the baryonic mass of the structure considered. The scatter around a relation of the type:
\begin{equation}\label{tully_fisher}
m \propto v^{\alpha}
\end{equation}
is surprisingly low, and seems to indicate that either there is no dark matter, or that there is a strict correlation between the dark matter and the baryonic matter component, which seems difficult to justify in the \LCDM model. MOND, on the other hand, predicts such a strict correlation and an exponent $\alpha = 4$ \citep{Famaey_McGaugh_2012}.

Let us now show that such power-law relations are a natural consequence of the D-M cosmology. We have seen that, due to the presence of the antimatter component on the outskirts of galaxies and clusters of galaxies, rotation curves are almost flat beyond a characteristic distance, which, to a good approximation, is about half the size of the depletion zone (see Fig. \ref{fig:periodic_dm_galaxy}), and about $2.5 r_v$, where $r_v$ is the virial radius as usually defined in the \LCDM cosmology.
At  distances $r > r_d/2 \approx 2.5 r_v$ from a condensed object, the D-M rotational velocity is predicted to be almost constant and equal to $v_d = \sqrt{2Gm/r_d}$.

The effective radius of the depletion zone\,\footnotemark[5] is in turn related to the mass $m$ of the structure considered, at least on average, and will be approximated here by the following simple equation between $m$ and $r_d$:
$m \approx (4/3) \pi r_{d}^3$ $ \overline{\rho}$
where $\overline{\rho}$ is the average density of matter.
Replacing this in the  above expression for $v_d$,  one finds that the asymptotic and nearly constant velocity is approximately proportional to  $r_d$, and the following approximate relation holds:
\begin{equation}\label{tully_fisher_dm}
m \propto v_d^3
\end{equation}

\footnotetext[5]{As noted previously, the depletion zones appear in fact more as a set of interconnected tubes
rather than isolated spheres.}

In this very crude approximation, where we have assumed spherical symmetry, the exponent of the TFR preferred in D-M is $\approx 3$, significantly smaller than the index 4 predicted by MOND (see e.g. \citealp{McGaugh_2000, Lelli_BTFR_2019}). On the other hand, we note that while the analysis of \cite{Lelli_BTFR_2019} finds a range of exponents between 3.5 and 4, with a preferred value of $3.85 \pm 0.09$, the  analysis of \cite{Ponomareva_2017} finds values of the exponent close to 3.0, with slope errors between 0.11 and 0.31 (see e.g. table 4 and figure 12 of \citealp{Ponomareva_2017}).

 It can also be noted that the estimate of the exponent $\alpha$ in the TFR relation varies rather widely depending on the photometric band used to infer the mass, although the near-infrared band using Spitzer photometry\,\citep{Lelli_BTFR_2019} may help to reduce this uncertainty.

\begin{figure}[]
 \begin{center}
\includegraphics[width=0.9\linewidth]{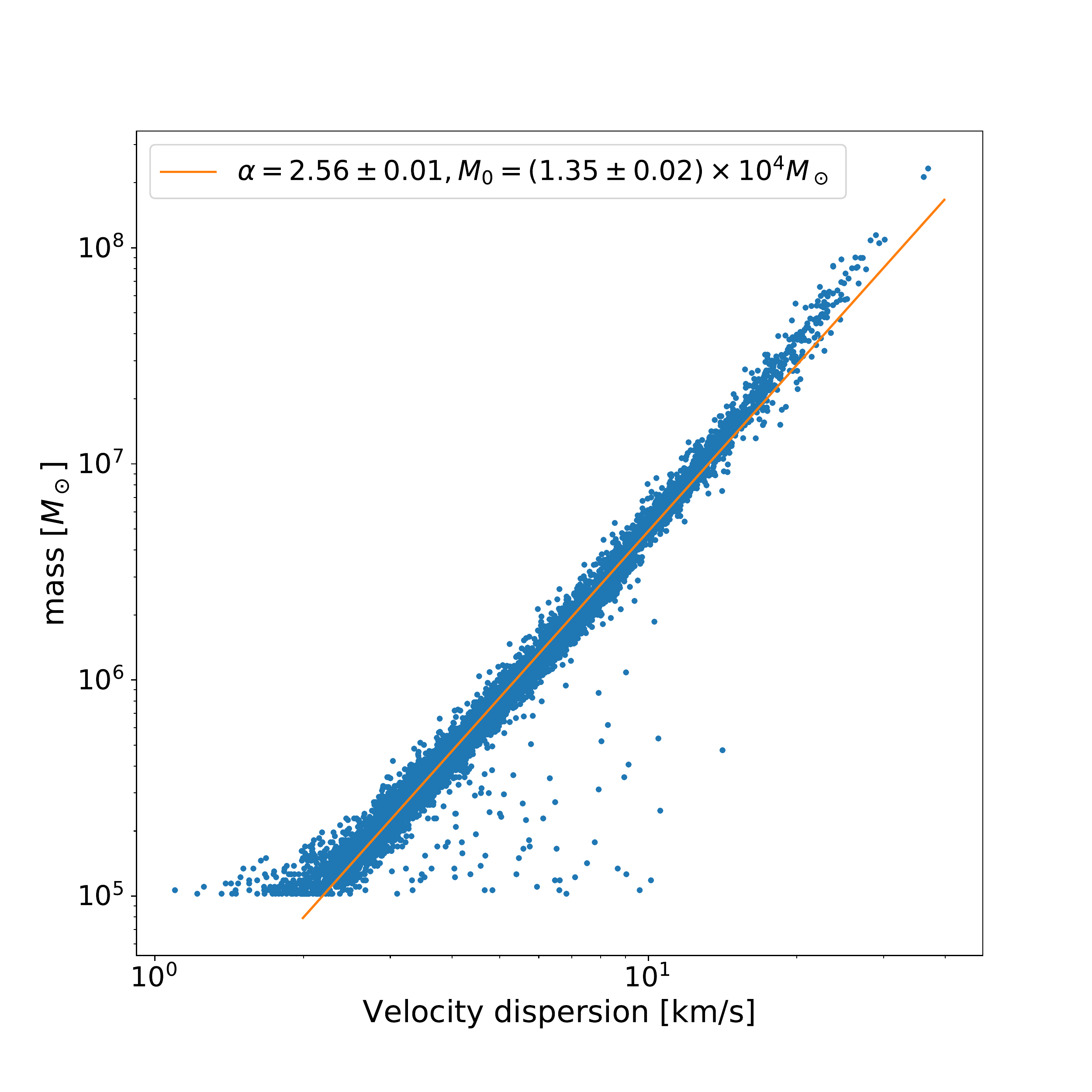}
 \end{center}
    \caption{FJR for halos with mass between $\approx 10^5$ and $\approx 2 \times 10^8 M_\odot$ in the D-M universe, using simulations of a modified version of the \texttt{RAMSES} code. The  mass versus virial velocity relation exhibits a power-law behavior with a very small scatter. The exponent in the power-law relation
    $m  \propto v^{\alpha}$ is $\approx 2.6$ when interpolating across the whole mass range, with $\alpha \approx 2.45$ and $\approx 2.85$ in the lower and the upper part of the plot, respectively. This behavior is similar to the power-law exponent of 3 favored by \LCDM, but somewhat smaller than the exponent 4 predicted by MOND.}
    \label{fig:Faber_Jackson_Dirac_Milne}
\end{figure}

In order to take into account more realistic situations than the idealized ``spherical" galaxy studied above, we have first analyzed the  FJR \citep{Faber_Jackson_1976}, using the \texttt{RAMSES} simulation described in the preceding section. Using the set of 7769 halos identified
by the AdaptaHOP algorithm\,\citep{Aubert_et_al_2004}, we calculated the velocity dispersion within the virial radius. We plot this quantity as a function of the halo mass within the same radius. The resulting scatter plot is shown in Fig.\,\ref{fig:Faber_Jackson_Dirac_Milne}, where the scatter beyond the power-law relation appears to be very small.

We consider this first analysis to be very encouraging, but we note that the FJR can be obtained also in the context of \LCDM, with a similar exponent. An important difference lies in the fact that in D-M, there is no dark matter, and that therefore the scaling relation between baryonic mass and virial velocity is tighter. On the other hand, we clearly need more realistic simulations, in particular regarding hydrodynamics and feedback, which will allow us to test the TFR for spiral galaxies, in addition to the FJR.

\section{MOND-like behavior in the D-M universe}\label{sec:MOND}

As mentioned previously, the MOND phenomenology can be justified if there exists gravitational polarization \citep{Blanchet_2007, Blanchet_LeTiec_2009}, while D-M predicts such a polarization. We have indeed seen in Section\,\ref{sec:simulations} that the D-M universe exhibits a MOND-like behavior for idealized, ``spherical" galaxies. On the other hand, the question remains open whether the value of the additional field created by antimatter and predicted by D-M can justify the characteristic acceleration $a_0 \approx 1.2 \times 10^{-10}$ m/s$^2$ that seems to best fit the MOND behavior. In the present section, we study the value of this additional gravitational acceleration and show that this value cannot be reduced in D-M to a constant and that, in particular, it depends on the redshift at which it is measured.

In order to study the intensity of the additional gravitational field created by the antimatter clouds on the outskirts of galaxies and clusters, we will compare
the modulus $|\vec{g}_m|$ of the gravitational field produced by matter alone given in equation \eqref{eq:gmatter} to $|\vec{g}_+|$, the total gravitational field acting on matter, and produced both by matter and antimatter as in equation \eqref{eq:g+}. Similarly, we will use the gravitational field produced by antimatter alone $|\vec{g}_{am}|$ to quantify the additional gravitational field present in D-M, due to the gravitational polarization between matter and antimatter, mimicking a MOND acceleration transition.

\begin{figure}[]
 \begin{center}
\includegraphics[width=\linewidth]{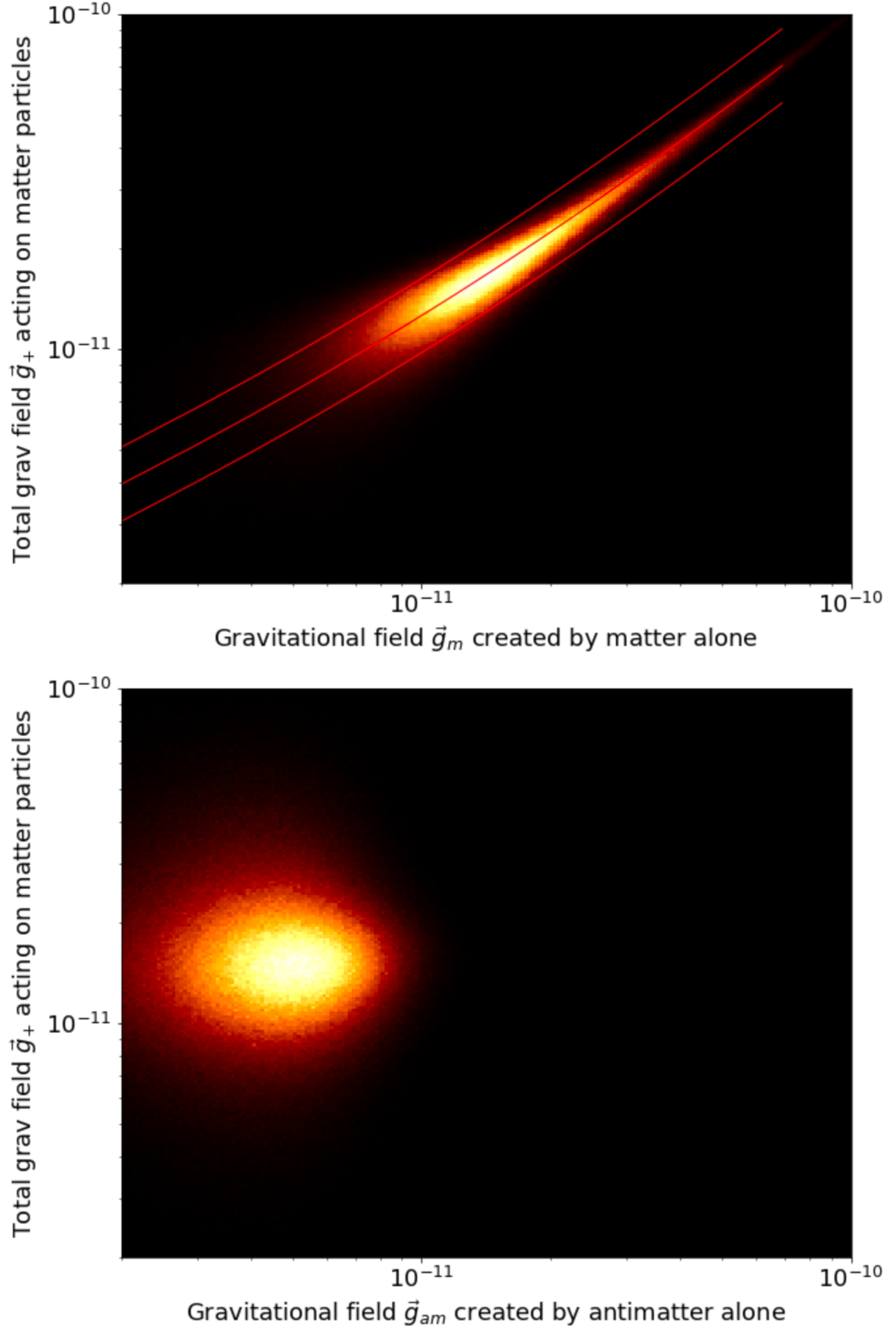}
\end{center}
    \caption{Top panel: scatter diagram showing the relation between the gravitational field $|\vec{g}_{m}|$ that would be created by matter if it were alone, on the $x$-axis, as a function of the total gravitational field $|\vec{g}_+|$ acting on a matter particle, created by both matter and antimatter, on the $y$-axis. While the high field part on the right part of the figure shows the expected Newtonian behavior, it can be seen that for accelerations smaller than $\approx 3 \times 10^{-11}$ m/s$^2$, a non-Newtonian behavior appears, with a flattening of the acceleration observed, in a behavior analogous to the MOND behavior.  The MOND interpolating function used in \cite{Lelli_BTFR_2019} with a value of the $a_0$ parameter equal to 0.4 $\times 10^{-11}$ m/s$^2$ has been superimposed on the simulation data, and the two adjacent curves represent the 1-$\sigma$ error of 0.11 dex found by these authors. Bottom panel: Scatter diagram showing the relation between the gravitational field $|\vec{g}_{am}|$ that would be created by antimatter if it were alone, on the $x$-axis, and the total gravitational field $|\vec{g}_+|$ acting on a matter particle, created by both matter and antimatter, on the $y$-axis. It can be seen on this figure that the gravitational field created by antimatter is much more uniform, with values smaller by typically one order of magnitude, than the Newtonian regime of the previous figure, created for the most part by matter.}
    \label{fig:Matter_Antimatter_DM_vs_matter_Newt}
\end{figure}

In the top panel of Fig.\,\ref{fig:Matter_Antimatter_DM_vs_matter_Newt}, the total gravitational field $|\vec{g}_{+}|$  acting on a matter particle, created by both matter and antimatter, is plotted as a function of the gravitational field $|\vec{g}_m|$ that would be created by matter if it were alone. These gravitational fields are calculated on a Cartesian $256^3$ grid (about 16 million cells), and the quantities estimated on the matter region (occupying $ \approx 50 \%$ of the grid volume), which is the only cut applied for the selection. The figure clearly shows two regimes: the high field part on the right part of the figure shows the Newtonian regime, where the gravitational field $|\vec{g}_{+}|$ matches almost exactly the expected Newtonian gravitational field $|\vec{g}_{m}|$ created by matter. It can be seen that for accelerations smaller than $\approx 3 \times 10^{-11}$ m/s$^2$, a non-Newtonian behavior appears, with a flattening of the observed acceleration, similar to the MOND behavior (but, on this figure, for a different value of the acceleration parameter $a_0$). Although the dispersion increases at low accelerations, it can be seen that, in this situation where the halo mass range exceeds three orders of magnitude, the vast majority of the points lie above the diagonal, indicating that the antimatter field $|\vec{g}_{am}|$ reinforces quite generally the matter field $|\vec{g}_m|$, mimicking a MOND behavior.

In the bottom panel of Fig.\,\ref{fig:Matter_Antimatter_DM_vs_matter_Newt}, we represent the modulus of the total gravitational field $|\vec{g}_{+}|$ acting on a matter particle, created by both matter and antimatter, as a function of the gravitational field $|\vec{g}_{am}|$ that would be created by antimatter if it were alone.
We can see that the additional acceleration created by antimatter in the matter and depletion zones is significantly smaller on average than the matter gravitational field observed in the Newtonian regime, and has a relatively peaked distribution, at $\approx 5 \times 10^{-12}$ m/s$^2$ in the cluster simulation presented.

Clearly, the MOND-like transition observed in Fig.\,\ref{fig:Matter_Antimatter_DM_vs_matter_Newt} at an acceleration of $\approx 3 \times 10^{-11}$ m/s$^2$ is here significantly smaller than the value of $\approx 1.2 \times 10^{-10}$ m/s$^2$ of the fundamental acceleration $a_0$ favored by MOND. Therefore, one may ask whether the D-M matter-antimatter scenario has any relevance as a possible explanation of MOND.
In order to answer this question, we note that the average modulus of the gravitational field created by antimatter depends on the redshift and on the largest scale of structure at this redshift. Fundamentally, the additional confining gravitational field created by antimatter, acting most of the time coherently with the matter field
(see Fig.\,\ref{fig:Matter_Antimatter_DM_vs_matter_Newt}), is almost independent of the mass of the individual galaxy considered, and equal to the antimatter field created at the scale of the largest structures.

\begin{figure}[]
 \begin{center}
\includegraphics[width=\linewidth]{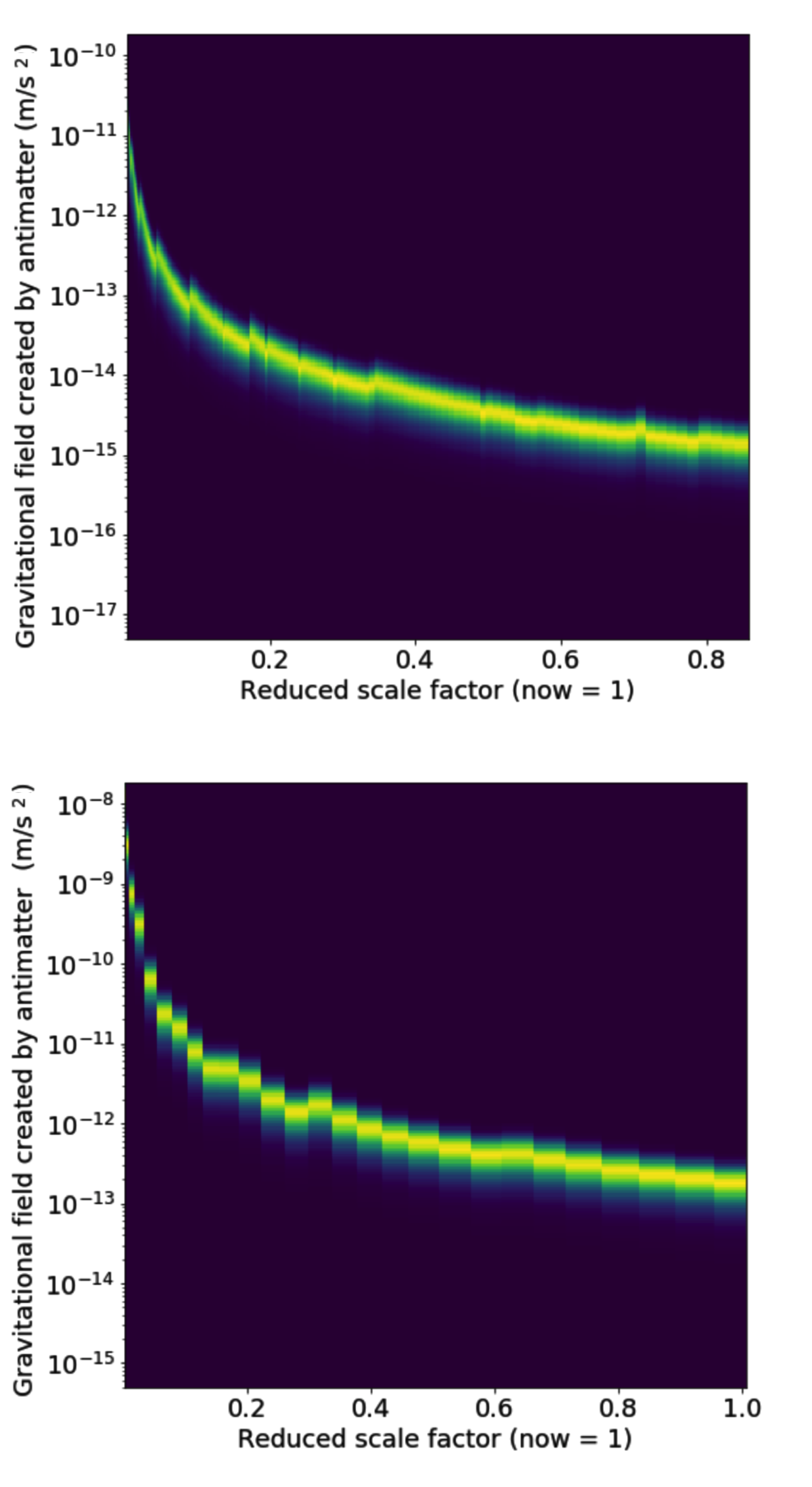}
 \end{center}
    \caption{Evolution as a function of the scale factor $a(t)$ of the distribution of the modulus of the gravitational field $|\vec{g}_{am}|$ created by antimatter for two simulations. In the first simulation, on the top panel, the total mass of the simulation is of galactic size, $\approx 2.2 \times 10^{10}$ M$_{\odot}$. The second simulation, on the bottom panel, is of cluster size and has a total mass of $\approx 10^{16}$ M$_{\odot}$.  Both distributions are rather peaked at all redshifts, and with an overall shape and width largely independent of redshift. On the other hand, the numerical value of the peak of the distribution clearly depends on the redshift. For both simulations, of limited size and mass, the peak value of the antimatter field differs from the fundamental constant postulated by MOND, $a_0 \approx 1.2 \times 10^{-10}$\,m/s$^2$. }
    \label{fig:Antimatter_grav_distr_waterfall}
\end{figure}

In order to emphasize this fundamental aspect, we have plotted in Fig.\,\ref{fig:Antimatter_grav_distr_waterfall} the evolution as a function of the scale factor $a(t)$ of the distribution of the gravitational field $|\vec{g}_{am}|$ created by antimatter for two simulations, the first one of ``galactic" size and total matter mass of about $2.2 \times 10^{10} M_\odot$, the second one of ``cluster" size and total matter mass of about $10^{16} M_\odot$.
These two figures present the following features: the distribution of the antimatter additional field is rather peaked at all redshifts. But although its overall shape is almost independent of redshift, reflecting the hierarchical build-up of gravitational structures, it is clear that the value of the peak depends not only on the redshift, but also, surprisingly, on the size of the simulation box, as long as it is smaller than a few homogeneity scales.

The latter implies that the value of the extra gravitational field due to antimatter in our Universe, much higher than in these two simulations, is still not numerically converged. We are currently designing a larger simulation with a $4096^3$ resolution and linear comoving dimensions of $\approx 1$ Gpc, which is expected to sample the homogeneity scale in D-M, and therefore to estimate more precisely the antimatter confining field predicted by D-M in our Universe. Ideally, the simulation volume would extend over a distance of a few Gpc, which is inaccessible at present.

Using the non-linear structure formation of D-M as a function of redshift shown in Fig. 4 of \cite{Manfredi2020}, we provide at this stage an estimate of the evolution of the average modulus $|\vec{g}_{am}|$ of the gravitational field created by antimatter as a function of redshift for a simulation of a region of a D-M universe of cosmological dimensions, \emph{i.e.} extending much beyond the homogeneity scale of $\approx 200$\,Mpc.

The quantity:
\begin{equation}\label{gam_evolution}
|\vec{g}_{am}| \approx \frac{4\pi G 2\rho(z)}{3} \frac{\pi} {k_{peak}(z)}
\end{equation}
where $\rho(z)$ is the average density at redshift $z$, and $\pi/k_{peak}(z)$ is the scale of the largest structures at redshift $z$, provides an estimate of the gravitational field created by antimatter at this redshift $z$.
It should be noted that the density contrast between matter and antimatter regions is of the order of unity, and for this reason $\rho$ enters directly in the above formula.

\begin{figure}[]
 \begin{center}
\includegraphics[width=\linewidth]{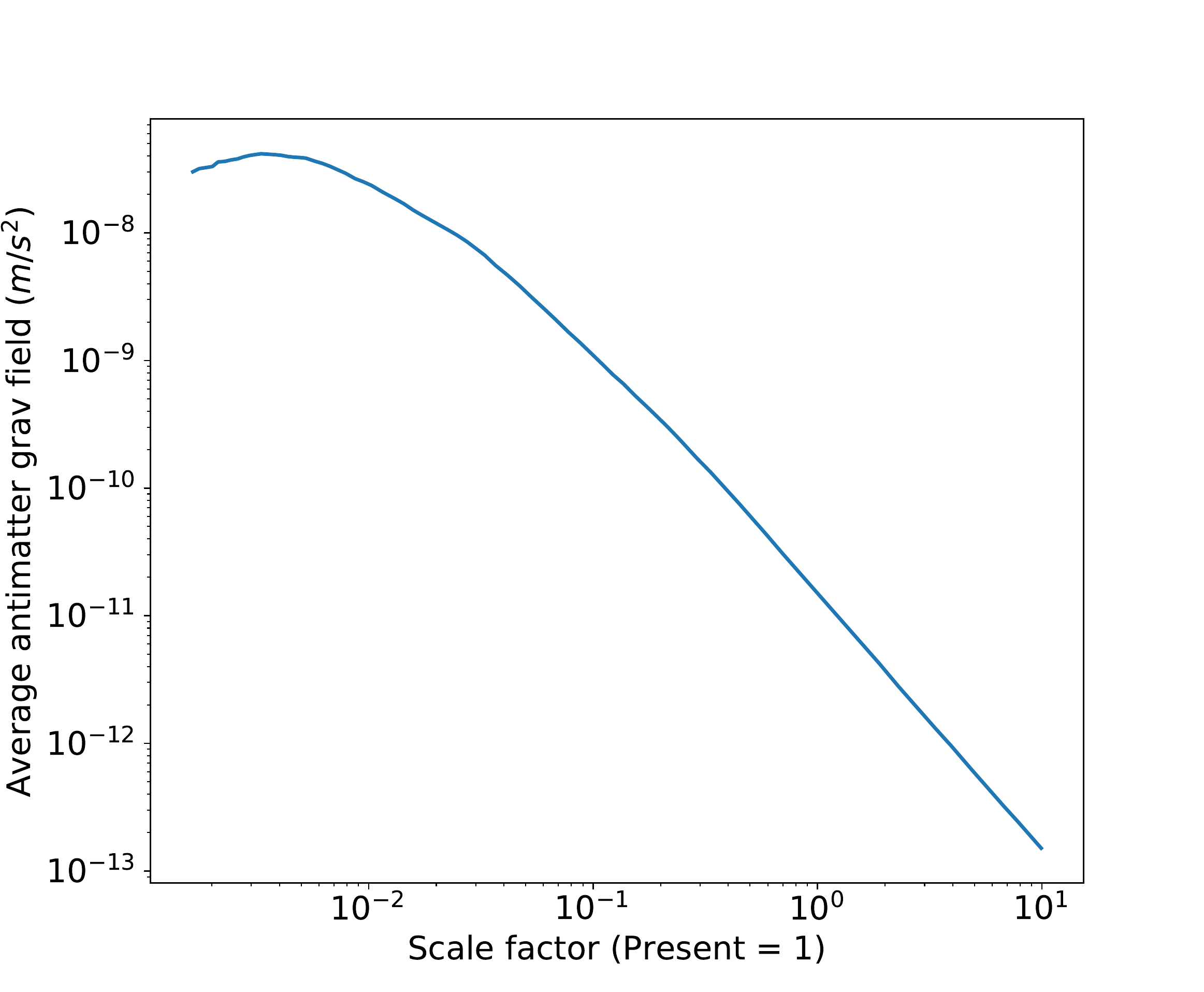}
 \end{center}
    \caption{Evolution as a function of the scale factor $a(t)$ of the average modulus of the gravitational field created by antimatter in the D-M universe. The scale factor is normalized such that at the present epoch the scale factor $a = 1$. At our epoch, the value of the average modulus of the antimatter gravitational field is of a few $10^{-11}$ m/s$^2$, \emph{i.e.} leading to a value of the $a_0$ acceleration parameter similar to that postulated by MOND. }
    \label{fig:Antimatter_grav_field_evolution}
\end{figure}

The evolution of this quantity is represented in Fig.\,\ref{fig:Antimatter_grav_field_evolution}. This figure shows two striking features: the modulus of the additional gravitational field created by antimatter is of the order of $2 \times 10^{-11}$ m/s$^2$ at the present epoch ($z \approx 0$). The conversion of this antimatter gravitational field to the $a_0$ parameter of MOND (using the parametrisations of \cite{Famaey_McGaugh_2012}) and the present uncertainty on the size of the initial domains of matter and antimatter results in a range of values of $a_0$ between $\approx 4 \times 10^{-11}$ m/s$^2$ and $\approx 2 \times 10^{-10}$ m/s$^2$, a rather striking similarity with MOND. It should be noted that this quantity clearly depends on the redshift, with a decreasing trend in the future, reflecting the fact that the hierarchical formation of large-scale structures has almost completely stopped at the present epoch in the D-M universe, while the adiabatic expansion continues. The prediction on the variation of this transition acceleration, which can be extracted from Fig.\,\ref{fig:Antimatter_grav_field_evolution}, might be used in observations of high-redshift galaxies to discriminate between the MOND and D-M scenarios. More precisely, once averaged over the range $a = [0.1, 1.0]$, the scale dependence of the antimatter gravitational field seems well approximated by a power-law $g_{am} \propto a^{-1.8}$. This might be tested by rotation curve data at relatively high redshifts, for example using the data from \citet{Genzel_Nature_2017}.

\section{Conclusions and perspectives}\label{sec:conclusions}
The present study has evidenced new elements of concordance between our Universe and the D-M universe. In particular, it proposes an explanation for the observation of flat rotation curves in galaxies, which is usually attributed to the presence of dark matter or to a modification of the laws of gravitation akin to MOND.  We have seen in particular that, due to the combined influence of the depletion zone and the antimatter clouds, flat rotation curves are generic in the D-M universe, leading to a systematic overestimate of the mass present in galaxies and clusters beyond a few ($\gtrapprox 2.5$) virial radii.

The additional force experienced in D-M differs both from the interpretation of MOND, with its modified expression for the gravitational field using a fundamental acceleration constant $a_0$, or from the local expression of the $\Lambda$ cosmological term conjectured by \cite{Gurzadyan1985} and \cite{Gurzadyan_Stepanian_2019}. On the other hand, in the same spirit as Gurzadyan, we propose a common explanation to the tentative dark energy and dark matter components of \LCDM, using in D-M a single constant $G$, instead of two constants $G$ and $\Lambda$. This provides an explanation for the otherwise rather extraordinary and fine-tuned coincidence of the dark energy and dark matter densities by linking them to the dynamical evolution of the matter and antimatter components, leading in D-M to a coasting universe, with $a(t) \propto t$\,(cf. Eq.\eqref{linear_expansion}).

We have also noted that the gravitational polarization between the positive and negative mass components \citep{Price_ajp1993} is at the origin of a MOND-like behavior. \cite{Blanchet_2007} and \cite{Blanchet_LeTiec_2009} have indeed shown that gravitational polarization could explain the MOND phenomenology, although these authors did not have in mind that this gravitational polarization could be due to antimatter. On the other hand, our analysis differs significantly from that of  \cite{Hajdukovic2011, Hajdukovic2014, Hajdukovic_2020}, which seems to lead to unobserved effects in the Solar system\,\citep{Banik_Kroupa_2020} and from the analysis by \cite{Penner2016}, who conjectured that MOND could be justified by the gravitational polarization of the vacuum, 
\emph{i.e.} without taking into account the gravitational polarization of matter and antimatter structures.

Additionally, using both a simple analytical model and \texttt{RAMSES} simulations, we showed that the D-M cosmology predicts a rather well-defined power law between mass on the one hand, and rotation or virialized velocity, on the other, in a large mass range of gravitational structures. This may provide an explanation for the impressive correlation evidenced in the TFR and FJR in galaxies and clusters,
although the exponent $\alpha$ in the relation $ m \propto v^{\alpha}$ was shown to be closer to 3, rather smaller than the preferred value of $3.85 \pm 0.09$ found by \cite{Lelli_BTFR_2019}, and the value of 4 predicted by MOND (see also \cite{McGaugh_BTFR_2012} for an earlier work focusing on gas-rich galaxies).

In future studies, we intend to realize a more realistic treatment of structure formation by including hydrodynamics and feedback in our \texttt{RAMSES} simulations, that will allow us to study the TFR, in addition to the FJR. We intend also to study larger simulation volumes, extending beyond the homogeneity scale ($\ga 200$ Mpc) predicted by D-M, and observed in our Universe (see, however, \cite{Keenan_KBC_2013} and \cite{Haslbauer_2020} for an example of inhomogeneity extending somewhat beyond this scale). Such large-scale simulations are a challenge for the D-M cosmology, as the  ``domains'' of matter and antimatter assembled at decoupling, which initiate the formation of larger structures, are of limited geometrical extension (of the order of 100 parsec at $z = 1080$, {\it i.e.} $\approx 100$ ckpc). It is important to note that the domains are in the non-linear regime (density contrast of order unity) almost immediately after the CMB transition, which we will study in a forthcoming publication. This requires a very high resolution compared to the usual cosmological simulations. These simulations at scales larger than a few hundred Mpc will allow us to test the deviations from local to large-scale measurements of the Hubble parameter $H_0$, a question of paramount interest in the present context, where the tensions on this parameter seem to reach or even exceed the $5 \sigma$ level between ``local" and cosmological measurements \citep{Riess2020, DiValentino_2021}.

Concerning the average value of the additional gravitational field created by the antimatter clouds, we have shown that the estimate of the transition acceleration $a_0$ at our epoch ($z = 0$) is of the order of $10^{-10}$ m/s$^2$, a striking similarity with the MOND formalism. Fundamentally, we note that the value of this additional ``MOND-like" gravitational field is determined by the antimatter field created at the largest structure scale, of the order of 200 Mpc. Also, we have shown that the modulus of this parameter depends on the redshift and is therefore not a fundamental constant, differing fundamentally from the MOND formalism.

Finally, we note that both nucleosynthesis and the almost purely non-linear structure formation in the D-M universe \citep{Benoitlevy, Manfredi2018, Manfredi2020} set  strong constraints on the size of initial ``domains" of matter and on the later hierarchical (largely bottom-up) development of structures. A complementary way to test our hypothesis will be to predict the mass distribution of stars and black holes resulting from the very early collapse of such matter domains. Compared to the black hole mass distribution derived by the LIGO and Virgo collaborations\,\citep{LIGO_VIRGO_Catalog_2020}, now in possession of about 70 candidate events (mostly binary black holes), this could represent an important additional test of the D-M scenario.

\begin{acknowledgements}
We are indebted to the anonymous referee and to Benoit Famaey, James Rich and Yves Sacquin for their thorough reading of the manuscript and their insightful comments.
Needless to say, they are not responsible for the errors and approximations remaining in this paper.
This work has made use of the Horizon Cluster hosted by the Institut d'Astrophysique de Paris.
The work of the YT\,\citep{Turk_2010}, IPython\,\citep{IPython_2007},
Matplotlib\,\citep{Matplotlib_2007}, NumPy\,\citep{NumPy_2011} and SciPy\,\citep{SciPy_2020}
development teams is also gratefully acknowledged.
The halo catalogs have been computed using the AdaptaHOP algorithm\,\citep{Aubert_et_al_2004}.
\end{acknowledgements}

\bibliographystyle{aa} 
\bibliography{rotation_curves_biblio}

\balance
\appendix
\section{Are there external field effects (EFE) in the D-M cosmology ?}\label{app:A}

The previous discussion on the rotation curves and the virial velocity distributions in the D-M universe has evidenced the surprising property that in this cosmology, to a good approximation, around a central massive object, a harmonic restoring force is felt in addition to the usual $Gm/r^2$ Newtonian force, leading to nearly flat rotation curves. At distances comparable to the radius of the depletion zone, this additional harmonic restoring force is far from negligible since it is on average equal to the force of the central galaxy (see Fig.\,\ref{fig:Sphere_grav_field}).
It may seem that this violates blatantly the shell theorem of \cite{Newton_1760} and, in General Relativity, the Birkhoff theorem\,\citep{Birkhoff_1923}.
In the present appendix, we discuss this question, which
presents interesting features.

The first element of answer comes from noticing that, in order to describe the gravitational behavior of matter and antimatter in the D-M universe, two coupled Poisson equations are required:
\begin{eqnarray}
\nabla^2\phi_{+} &=& 4\pi G (\rho_{+} - \rho_{-}), \\
\nabla^2\phi_{-} &=& 4\pi G (-\rho_{+} - \rho_{-}) \ .
\end{eqnarray}

The possibility to express the D-M gravitational behavior with coupled equations using Laplacian operators implies, through the Gauss theorem, that in a situation of spherical symmetry, the gravitational field in an empty region must be zero, since the mass enclosed is zero and spacetime in this region should be Minkowskian. However, there is a caveat to this assertion, related to the boundary conditions of the mass configuration.
The previous statement on the shell theorem being respected may indeed seem at odds with our decomposition in terms of three spheres (see Fig.\,\ref{fig:Sum_3_cubes}) of the average galaxy environment in the D-M universe, where a harmonic restoring force is observed in addition to the usual Newtonian force within the depletion zone. The short answer to this apparent contradiction is that the situation lacks spherical symmetry and that the harmonic restoring force that we have derived just approximates this asymmetric configuration of the antimatter cloud surrounding a spherical depletion zone (see in particular Fig.\,\ref{fig:periodic_dm_galaxy}). Note in particular that in actual configurations, the depletion zones are not spherical, but percolate with surrounding depletion zones, with a similar percolation property for the antimatter clouds. Also, the approximation used is valid only at distances smaller than $r_d$, where $r_d$ is the approximate size of the depletion zone surrounding the massive structure, and becomes increasingly inaccurate when we exceed radial distances larger than $\approx r_d/2$.

Although this asymmetry correctly answers the question on the violation of the shell theorem in galactic configurations in the D-M universe, it may hide some interesting elements of discussion, which we summarize in the following.

The first element comes from considering the situation resulting from the superposition of only the two cubes (a) and (c) of Figure\,\ref{fig:Sum_3_cubes}, \emph{i.e.} the cube with uniform repulsive background and the cube containing a sphere with uniform positive mass density compensating, within the volume of the sphere, the negative mass background of the first cube. As soon as we have accepted the property that the gravitational field created by cube (a) with uniform density is necessarily zero everywhere, which seems unavoidable by symmetry, it is also clear that the contribution of the second cube will create a harmonic restoring force, although the inner sphere, in the superposition of the two cubes, is now empty.

The situation is even stranger when we consider the configuration with complete spherical symmetry, where the whole space is filled with a uniform negative background, to which we superimpose a sphere centered on the origin with a positive uniform density, compensating the negative mass fluid inside the sphere (and only there). This time we cannot invoke the asymmetry of the situation and it seems that we have a gross violation of the shell and Birkhoff theorems\,\citep{Newton_1760, Birkhoff_1923} since the gravitational field appears to be nonzero in an empty region with exact spherically symmetry.

However, the expression of Birkhoff's theorem\,\citep{Birkhoff_1923} only states that any spherically symmetric solution of the vacuum field equations must be static and asymptotically flat, and represented by a Schwarzschild metric. We must then note that we have in fact filled out the entire space with a negative mass fluid of constant density, and therefore with infinite negative global mass. This configuration is necessarily not static, so the ``cosmological" aspects must now be taken into account in the dynamical situation. Our decomposition into three cubes with respective masses $-2m$, $+m$ and $+m$, on the other hand, restores a total mass zero, and has not a diverging mass and potential at infinity\,\citep{Seeliger_1895}, but involves a configuration without spherical symmetry.

Two important additional comments can be made:
\begin{itemize}

\item{ The first comment is based on the remarkable analysis by Gurzadyan as early as 1985 \citep{Gurzadyan1985},
in the early days of dark matter searches, which in several respects reaches conclusions similar to those of the present analysis\,\footnotemark[6].
\footnotetext[6]{The last sentence of this paper is particularly noticeable: ``The smallness of the cosmological constant evidently excludes the checking of Equation (5) by means of any experimental methods; however, the contribution of the second member in (5) can be evaluated from the analysis of the structure of galaxy clusters, their haloes, etc. The possibility of the existence of a long-range force of the above type may affect in a certain way the ideas concerning the future of the open Universe".}

In his first publication on this topic\,\citep{Gurzadyan1985}, Gurzadyan notes that the Newtonian $1/r^2$ force law could be extended by requiring the property,
realized in the Newtonian case, that for a configuration with spherical symmetry, the gravitational action of a mass can be reduced to the situation where all the mass is concentrated at the origin. Remarkably, this requirement leads not only to the Newtonian potential, but also to an additional harmonic force,
attractive or repulsive depending on the sign of $\Lambda$, which appears as the fluid analog of the cosmological constant, and we reproduce here, with a slight change of notation, Equation (5) of \cite{Gurzadyan1985}:
\begin{equation}\label{gurzadyan}
F(r) = Ar^{-2} + \Lambda r
\end{equation}
In this sense, the introduction of a cosmological constant in Newtonian cosmology finds a natural justification in the Milne cosmology \citep{Milne}.
Gurzadyan notes that this second component has the property that the shell theorem (zero gravitational field inside an empty spherical shell) is {\it not} respected.
More importantly, in recent publications, \citep[see {\it e.g.}][and references therein]{Gurzadyan_Stepanian_2019},
Gurzadyan conjectures that the flat rotation curves observed in galaxies can be explained by the same (local) expression as the cosmological repulsive term. However, with his expression of a generalized gravitational potential
using {\it two} gravitational constants $G$ and $\Lambda$,
Gurzadyan does not consider the possibility that these two terms could be expressed with a {\it single} constant $G$ but with a sign
reversal. For example, the potential term in the GR metric
in equation (1) of \citet{Gurzadyan_Stepanian_2019} leads directly in Equation (2) to the two components,
Newtonian and harmonic restoring force, of the gravitational field that we have derived previously for D-M.\newline}
\item{The second comment is related to the fact that, as mentioned previously, cosmological simulations of self-gravitating structures routinely use effective negative mass without explicitly stating it.  Indeed, as mentioned previously, cosmological simulations first calculate the average density in order to derive the average cosmological expansion of $a(t)$, the cosmological scale parameter. After subtraction of this average density, a symmetric (in terms of sign of mass) mass distribution is then obtained,  usually with symmetrical overdensity and underdensity distributions. Piran indeed remarked \citep{Piran1997}, following previous numerical simulations of gravitational structures with Dubinski and collaborators (\citealt{Dubinski1993}), that while we are used to representing gravitational structures in terms of (collapsing) positive mass, it is also possible and useful to consider the problem in terms of (expanding) voids of negative mass. In their expansion, these voids will effectively develop into the largest structures in the universe. As studied later by \cite{sheth2004}, now followed by the work of several authors
(for a review, see for example \citealt{Pisani_void_review_2019} and references therein), the study of voids constitutes today a powerful test of the cosmology at play in our Universe. Indeed, the characteristics of 
the KBC void\,\citep{Keenan_KBC_2013} appear to rule out the \LCDM cosmology at the $6 \sigma$ level\,\citep{Haslbauer_2020}.

We further note, following Piran, that the matter surrounding an underdense region will create in its expansion a high density ridge
along the rim of the underdense region, and density even diverging at shell crossing (see e.g. Figs.\,1 and 2 of \citep{Dubinski1993}). A region nearly empty, or even totally empty, will therefore, in a cosmological context,
effectively ``repel" the surrounding matter, due to its faster expansion compared to its surroundings. Indeed, the Hubble tension might be resolved by taking into account the outflow from the KBC void \citep{Haslbauer_2020}.

Piran goes as far as to suppose that there could exist
a species of matter with negative gravitational mass and positive inertial mass, violating the usual expression of the Weak Equivalence Principle. But curiously, Piran describes the interactions of this new species
endowed with negative gravitational mass and positive inertial mass as attractive between themselves.
We note that this is not the behavior actually observed in ``voids" as the negative density (underdensity) flattens out instead of becoming more negative. Of course, under the ordinary assumption that only positive mass particles exist, the maximum ``negative" mass (underdense) region is limited by the condition of zero matter density. This leads to the fundamental new element introduced by the introduction of ``negative" mass particles: a depletion zone develops, and we have seen the fundamental role that it plays in terms of mimicking the behavior of extended dark matter clouds or, alternatively, providing a MOND-like behavior.
As we have studied in detail in \cite{Manfredi2018}, if we want to describe the behavior of the repulsive underdense regions using a second species of ``negative" mass, this is not possible in a Newtonian description, even with the three parameters of inertial, passive gravitational and active gravitational mass.
In order to implement the gravitational behavior of the Dirac particle-hole system, let us stress again that it is necessary to use a bimetric description \citep{Manfredi2018}.}

\end{itemize}

In conclusion, the harmonic restoring force predicted by the D-M model, adding its contribution to the usual Newtonian force term, is an unavoidable consequence of the combined influence of the antimatter cloud and empty depletion zone developing in the D-M universe around a localized structure of positive mass.
But unlike the additional $\Lambda$ term conjectured by Gurzadyan, this harmonic restoring force does not violate
the shell theorem. This unexpected restoring force is simply explained in terms of the asymmetric distribution of the positive and negative mass
components in the D-M cosmology. 

\end{document}